\documentclass[prb,aps,amssymb,shownopacs,twocolumn,superscriptaddress,10pt]{revtex4}
\usepackage{amsmath}
\usepackage{amssymb}
\usepackage{amsthm}
\usepackage{amsfonts}
\usepackage{listings}
\usepackage{enumerate}
\usepackage{latexsym}
\usepackage{psfrag}
\usepackage{bm}
\usepackage[all]{xy}
\usepackage{graphicx}
\usepackage{subfigure}
\usepackage[pdftex,colorlinks=true]{hyperref}
\usepackage{xcolor}
\usepackage{H1}

\newcommand{\w}{\omega}
\newcommand{\fr}{\frac}

\newcommand{\mbb}{\mathbb}
\newcommand{\ii}{\mathrm{i}}
\newcommand{\lag}{\mathcal{L}}
\newcommand{\partition}{\mathcal{Z}}

\begin{document}
\title{Field Theories for Gauged Symmetry Protected Topological Phases: Abelian Gauge Theories with non-Abelian Quasiparticles}

\author{Huan He}
\affiliation{Physics Department, Princeton University, Princeton, New Jersey 08544, USA}

\author{Yunqin Zheng}
\affiliation{Physics Department, Princeton University, Princeton, New Jersey 08544, USA}

\author{Curt von Keyserlingk}
\affiliation{Princeton Center for Theoretical Science, Princeton University, Princeton, New Jersey 08544, USA}

\date{\today}

\begin{abstract}
Dijkgraaf-Witten (DW) theories are of recent interest to the condensed matter community, in part because they represent  topological phases of matter, but also because they characterize the response theory of certain symmetry protected topological (SPT) phases. However, as yet there has not been a comprehensive treatment of the spectra of these models in the field theoretic setting -- the goal of this work is to fill the gap in the literature, at least for a selection of DW models with abelian gauge groups but non-abelian topological order. As applications, various correlation functions and fusion rules of line operators are calculated. We discuss for example the appearance of non-abelian statistics in DW theories with abelian gauge groups.
\end{abstract}

\maketitle

\tableofcontents

\section{Introduction}
\label{sec: intro}

Landau-Ginzburg symmetry breaking formalism is one of the fundamental building blocks for conventional condensed matter physics\cite{landau1937theory,ginzburg1950theory,landau1958lifshitz}. Historically, it applies successfully to understanding magnets, BCS superconductor, and etc.
Within the Landau-Ginzburg paradigm, phases are characterized by their global symmetries and which of those symmetries are spontaneously broken. The result is that phases can be paramagnetic with short ranged correlations, or have long range correlations associated with spontaneously broken symmetries. More recently, it has been found that this characterization is a) too coarse, and b) needs to be extended. First, not all paramagnetic phases with a certain global symmetry group are identical. Indeed, there exist paramagnetic symmetry protected topological (SPT) phases which have a global symmetry which is not broken spontaneously \cite{chen2011classification,chen2010local,levin2012braiding,burnell2014exactly,chen2011complete,chen2011two,chen2012chiral,chen2012symmetry,chen2013symmetry,liu2011symmetry,gu2015multi,wang2015bosonic,wang2015field,wang2015non,ye2013symmetry,ye2015vortex,pollmann2012symmetry,fidkowski2010effects,gu2009tensor,wen2013classifying,wen2013topological,lu2012theory,haldane1983nonlinear,affleck1987critical,haldane1983continuum, vishwanath2013physics,fidkowski2011topological,pollmann2010entanglement}, but which nevertheless cannot be adiabatically connected to one another in the presence of the protecting global symmetry. Secondly, the Landau-Ginzburg paradigm does not account for  topologically ordered states\cite{WenNiu90,Wen90top2}, which are not characterized by the spontaneous breaking of global symmetries and the long distance correlations of local order parameters. 

The mentioned SPTs and topological phases both have a connection to topological quantum field theories (TQFTs). The low-energy long-distance behavior of topological phases is described by TQFTs. Bosonic SPTs on the other hand, can be characterized by gauging their protecting global symmetry as done in Ref.~\onlinecite{levin2012braiding}. After gauging the global symmetry, any SPT phase becomes  topologically ordered, and it turns out that distinct SPT phases become distinct topological orders when gauged. For example, there are two different SPTs with a $\mbb{Z}_2$ global symmetry in (2+1)D. After gauging the $\mbb{Z}_2$ symmetry of these two SPTs, one becomes deconfined $\mathbb{Z}_2$ gauge theory and the other becomes the distinct double semion topological order\cite{levin2012braiding}. The classification of a large class of SPTs then simply becomes the classification of different possible topological orders or TQFTs with gauge group $G$. When $G$ is unitary and finite, such TQFTs are called ``Dijkgraaf-Witten" (DW) models, which are classified by the cohomology group $H^{d+1}(G,U(1))$\cite{dijkgraaf1990topological}, where $G$ is the given on-site global symmetry and $d$ is the spatial dimension. Hence, bosonic SPTs with finite on-site global unitary symmetry group $G$ are classified by an element of $H^{d+1}(G,U(1))$ (called a ``cocycle"), and are in 1-to-1 correspondence with DW models. In addition to their connection to bosonic SPTs, DW models are themselves interesting examples of topological orders. Therefore by studying DW models we learn something about SPTs, and gain insight into a wide range of possible topological orders. 

We are thus motivated  to study DW field theories, and in this work we focus on (2+1)D. We further focus on DW theories with finite abelian gauge groups. Topological field theories with non-semisimple gauge groups were studied in Ref~\onlinecite{Ferrari2003,Ferrari2007,Ferrari2015}. On the one hand, the DW theories were originally introduced as field theories with topological terms directly related to the cocycle in question\cite{dijkgraaf1990topological}. Subsequently, the same cocycle data was found to encode an algebraic structure called a quasi-quantum double\cite{propitius1995topological}, which it was proposed should describe the algebra of anyon excitations in the DW model. The goal of the present work is to provide a clearer bridge between the field theory for the DW model, and the algebraic theory of its anyonic excitations. We do this by explicitly constructing line operators in the DW field theory, and calculating their braiding and fusion rules. 

Some DW theories (called type I and type II\cite{propitius1995topological})  can be thought of as continuum `K-matrix' Chern-Simons theories, and their line operators are already well understood\cite{lu2012theory,cheng2014topological}. However, the line operators in more exotic `type III' DW field theories have remained elusive (Field theories of type III DW model were also discussed in \onlinecite{PhysRevLett.112.231602,kapustin2014anomalies,wang2015bosonic,gu2015multi}). In this work we focus on the very simplest such theory -- the type III twisted  DW model with gauge group $\mbb{Z}_2^{\otimes 3}$. We construct all line operators (Wilson lines, flux insertion lines and their composites). Instead of the naive 64 line operators (8 Wilson operators, 8 flux insertion operators and their composites), we find that the number of distinct line operators is only 22, reproducing an algebraic result in Ref.~\onlinecite{propitius1995topological}.  Moreover, we can compute the correlation functions and fusion rules for these operators and confirm that the  type III twisted $\mbb{Z}_2^{\otimes 3}$ theory is a non-abelian topologically ordered phase. We also explain how these results extend to more general abelian gauge groups. Our results should not be considered as completely mathematically rigorous -- we use a continuum field theory formalism on the understanding that at certain key points in the calculation the lattice regularization needs to be considered carefully. In this manner, our approach is of a similar level of rigor to other continuum approaches used to understand similar models\cite{Ye:2015eba,kapustin2014coupling,gu2015multi}. The utility of our formalism is that it readily allows us to derive a number of non-trivial results in a reasonably intuitive manner, without the need for understanding some the more abstract algebraic machinery behind the existing group cohomology results.

The rest of the paper is organized as follows:  In Sec.~\ref{sec: 2+1D}, we briefly summarize the $K$-matrix formulation of type I,II cocycles. Then we attack the problem of type III cocycles. We begin by constructing all of the line operators for the type III twisted $\mbb{Z}_2^{\otimes 3}$ theory in (2+1)D and work out their correlation functions, and fusion rules. This theory turns out to be a non-abelian topologically ordered state. Sec.~\ref{sec. ZN examples} further generalizes these results to the type III twisted $\mathbb{Z}_N^{\otimes3}$ theory. We also provide three appendices for further details: App.~\ref{app. coefficient} explains how we fix the coefficients in the Lagrangians; App.~\ref{app. path integral} explains how to calculate the path integrals with a lattice regularization; App.~\ref{app.sec. gauge} varifies the gauge invariance of flux insertion operator $V_{n_1n_2n_3}$ on lattice explicitly;  App.~\ref{app. ModMat}, we provide the modular $S$, $T$ matrices for the type III twisted $\mathbb{Z}_2^{\otimes 3}$ DW theory; App.~\ref{app. QD} shows quantum double calculations for type III $\mathbb{Z}_N^{\otimes 3}$ theory used to verify the field theoretic results from the main text.

\section{DW models in continuum formalism }
\label{sec: 2+1D}
DW theories were first formulated as lattice gauge theories\cite{dijkgraaf1990topological}. Consider a (2+1)D theory with abelian gauge group $G$. The DW theory action is encoded by some 3-cocycle $\omega: G\times G \times G \mapsto \text{U}(1)$. The DW action is obtained by  performing a simplicial decomposition of the manifold in question and orders the vertices, to write down a partition function weight
\begin{equation}
\prod_t \omega(\vec{A}^{01}_{t},\vec{A}^{12}_{t}, \vec{A}^{23}_{t}),
\end{equation}
where  $\vec{A}^{e}_{t}$ is a $G$-valued flat gauge field living on an edge $e$; $t$ are 3-simplices obtained by triangularizing the spacetime manifold; and $01$, $12$, $23$ are specific edges in $t$ determined by the ordering on the simplicial decomposition\footnote{Gauge invariance is ensured by the so-called cocycle condition on $\omega$ -- see Ref.~\onlinecite{dijkgraaf1990topological}.}. Note that in this construction, $\vec{A}$ is assumed to be flat; one goal in this work is to extend the construction above so as to relax this constraint on $\vec{A}$ whilst maintaining gauge invariance. This in turn allows us to examine the full spectrum of the DW model, and explicitly construct all the line operators in the theory. In addition, the lattice actions considered in  Ref.~\onlinecite{dijkgraaf1990topological} are difficult to work with. A second goal of this work is to formulate in detail a more convenient and transparent continuum version of these field theories much like those in Ref.~\onlinecite{kapustin2014coupling} -- it will turn out that the explicit regularization of the models is for many purposes unimportant.

Our discussion is organized as follows: in \secref{subsec: type I II} we briefly discuss the type I and type II cocycles which leads to abelian topological phases; in \secref{subsec: continuum} we construct the continuum action for type III DW models; in Sec. \ref{subsec: 22 line operators}, we argue that there are only 22 distinct line operators (in agreement with Ref.~\onlinecite{propitius1995topological}), instead of the naive 64 operators one expects in a $\mathbb{Z}^{\otimes 3}_2$  gauge theory. In Sec. \ref{subsec: fusion rules 2+1} and \ref{subsec: correlation 2+1}, we calculate correlation functions and fusion rules of these line operators. Further generalizations to $\mbb{Z}_N^{\otimes 3}$ can be found in Sec.~\ref{sec. ZN examples}.

\subsection{Type I and Type II Cocycles}\label{subsec: type I II}
Before we delve into the field theory for type III cocycle, let's briefly comment on the type I and type II cocycles. The cocycles of abelian discrete groups $\prod_i \mathbb{Z}_{k_i}$ have been categorized into three types\cite{propitius1995topological,wang2015field}. Type III cocycle is the focus of this paper and will be explained in the following texts. Here we only briefly discuss continuum field theories corresponding to the type I and type II cocycles, as they are just special cases of abelian $K$-matrix theories. 

As an example, for a discrete group in the form of $(\mathbb{Z}_{N})^{\otimes L}$, the type I and type II cocyles can be written uniformly as\cite{propitius1995topological}
\begin{equation}\label{eq. type I II}
\omega^{ij}(A,B,C) = \exp\left( \frac{2\pi\ii M^{ij}}{N^2} a^{i}(b^{j}+c^{j}-[b^j+c^j])\right),
\end{equation}
where $A, B, C \in \mathbb{Z}_{N}^{\otimes L}$; $a^i,b^i,c^i \in\{0,1\ldots,N-1\}$ for $i=1,2,\ldots L$ label the $i$-th component of these group elements in the $L$ copies of $\mathbb{Z}_{N}$ respectively; $M^{ij}$ are integers valued in $\{0,1,\ldots,N-1\}$. The bracket notation is defined by $[x]:= x \mod{N}$ with $[x]\in \{0,1,\ldots,N-1\}$

The continuum field theories corresponding to the type I and II cocycles Eq.~\eqref{eq. type I II}, have action  $\frac{1}{4\pi}\int K^{i j} a^i d a^j$ where $i\in1,\ldots ,2L$ and each $a^i$ is a compact $\text{U}(1)$ connection 1-form. The corresponding $K$-matrix is
\begin{equation}
\left(
\begin{matrix}
0	&	N \mathbb{I}_{L}	\\
N \mathbb{I}_{L}	&	M+M^T
\end{matrix}
\right),
\end{equation}
where $\mathbb{I}_{L}$ is $L$-by-$L$ identity matrix, $M$ is an $L$-by-$L$ integer matrix whose elements are just $M^{ij}$ in Eq.~\eqref{eq. type I II}. The  type I and type II theories hence only produce abelian topological order and all line operators and their statistics/correlations are well known\footnote{See Ref.~\onlinecite{Wen04,WenReview2016} for review articles, and references therein.}.

\subsection{Type III Cocyles}\label{subsec: continuum}
Having summarized the story for type I and II cocycles, we describe the so-called type III twisted DW theory. These are characterized by a 3-cocycle of form
\begin{equation}\label{eq:3cocycle}
\omega(A,B,C) = e^{2 \pi i p a^{1} b^{2} c^{3}/k_1 k_2 k_3},
\end{equation}
where $A, B, C \in \mathbb{Z}_{k_1}\times\mathbb{Z}_{k_2} \times\mathbb{Z}_{k_3}  $ and $a^i,b^i,c^i =0,1\ldots,k_i-1$ for $i=1,2,3$ label the components of these three group elements in the three copies of $\mathbb{Z}_{k_1},\mathbb{Z}_{k_2},\mathbb{Z}_{k_3}$ respectively.  Here, 
\begin{equation}
p=n k_1k_2k_3/\gcd(k_1,k_2,k_3),
\end{equation}
where $n \in \mathbb{Z}_{\gcd(k_1,k_2,k_3)}$ labels the distinct possible choices of cocycle. Using the above prescription, the DW models are rigorously formulated on the lattice. However, many of the known abelian examples of these theories are more conveniently formulated in the continuum. For instance Refs.~\onlinecite{tiwari2016wilson,kapustin2014coupling,wang2015field,ye2013symmetry,gu2015multi} characterize certain abelian DW topological orders in terms of continuum toy models. In this spirit, we start by writing down the most naive interpretation of the 3-cocycle \eqnref{eq:3cocycle} in the continuum and examine under which conditions it is gauge invariant. The Lagrangian for $\mbb{Z}_{k_1}\times \mbb{Z}_{k_2} \times \mbb{Z}_{k_3}$ theory is:
\begin{equation}\label{eq: lag 2+1}
\lag= \fr{k_i}{2\pi} b_i \wedge dA_i + \fr{p\epsilon^{ijk}}{6(2\pi)^2}A_i \wedge A_j \wedge A_k,
\end{equation}
where the repeated indices imply summation. We now clarify the above notation: $A_{i=1,2,3}$ are the components of $\vec{A}$ in $\mbb{Z}_{k_1}\times \mbb{Z}_{k_2} \times \mbb{Z}_{k_3}$, and where $i,j,k$ are summed over $\{1,2,3\}$. The first term is a $bF$ term which enforce the flatness condition of $A_i$ fields in the partition function, and the second term is the type III twist term. For $G=\mbb{Z}_2^{\otimes 3}$ we have $k_i=2~(i=1,2,3)$. In this case there are two possible choices for $p$: $p=0$ corresponds to plain $\mbb{Z}_2^{\otimes 3}$ gauge theory (3 copies of $\mathbb{Z}_2$ model), while  $p=4$  we  refer to as `twisted' $\mbb{Z}_2^{\otimes 3}$ gauge theory. Following previous work on these theories\cite{wang2015field}, we detail how to fix the possible values of the coefficients of the twist terms in \eqnref{eq: lag 2+1} in \appref{app. coefficient}.

As written, the action is invariant under transformations
\begin{equation}\label{eq: gauge transf 2+1}
\begin{split}
& b_i\rightarrow b_i+d\beta_i+\fr{p\epsilon^{ijk}}{2\pi k_i}(A_j\alpha_k-\fr{1}{2}\alpha_jd\alpha_k)	\\
& A_i\rightarrow A_i+d\alpha_i,\;i=1,2,3\;.
\end{split}
\end{equation}
where $\alpha,\beta$ is a scalar field and we have omitted wedge products for brevity\footnote{The noncommutative gauge transformations have indicated that the theory is actually non-Abelian. We thank E. Witten for pointing it out. Also see Ref.~\onlinecite{gu2015multi}.}. In addition both gauge fields are presumed to be compact insofar as
\begin{align}
b_i &\equiv b_i+ 2\pi \nonumber\\
A_i &\equiv A_i +2\pi\label{eq: comp transf 2+1}.
\end{align}
Here $b_i$ and $A_i$ are understood as the value of gauge field on a bond of spacetime lattice. 

As mentioned before, in the partition function,
\begin{eqnarray}
\partition:=\int D[A_i]D[b_i] \exp(\ii\int\lag)	\;,
\end{eqnarray}
$b_i$ fields play the role of Lagrangian multipliers and enforce the flatness constraint on $A_i$.  However, once $b$ sources are inserted in the path integral, $A_i$ fields are no longer flat. To see this, note that in the presence of $b$ sources, the path integral takes the form
\begin{equation}
\partition[\gamma]=\int D[A_i]D[b_i] \exp\db{\ii\int\lag+\ii\oint_{\gamma} b_i \ldots}.
\end{equation}
Once the $b_i$ fields are integrated out, $dA_i$ is enforced to be nonzero on $\gamma$ i.e., $\textstyle{dA_i=\frac{2\pi}{k_i}\gamma^{(2)}}$ where $\gamma^{(2)}$ is the 2-form Hodge dual to the contour $\gamma$. However, a single term, $\exp\left(\ii\oint_{\gamma} b_i\right)$, is not gauge invariant. Hence, it cannot be a valid operator for the twisted type III  $\mbb{Z}_2^{\otimes3}$ theory. We discuss the all valid line operators in the following section. And we coin the operators involving $\exp\left(\ii\oint_{\gamma} b_i\right)$ ``flux insertion operators" for the following text.

\subsection{Line Operators}
\label{subsec: line operators 2+1}
In this section, we construct all the Wilson operators and flux insertion operators on a given loop $\gamma$, for the type III twisted $\mbb{Z}_2^{\otimes 3}$ field theory mentioned above, Eq.~\eqref{eq: lag 2+1}. We adopt the notation $U_{pqr}$ for Wilson operators, and $V_{pqr}$ for flux insertion operators, where $p,q,r=0,1$. We will see that when $p,q,r\equiv 0 \mod 2 $, the resulting operators are trivial in the sense that they have trivial correlations with other operators.

An essential requirement of constructing these loop operators is that they are invariant under gauge transformation Eq.~\eqref{eq: gauge transf 2+1}. Moreover, the line operators should also be invariant under $A_i \mapsto A_i + 2\pi$ and $b_i \mapsto b_i + 2\pi$, because the gauge fields are assumed to be compact with $2\pi$ periodicity. Following the gauge invariance principle, the Wilson operators can be written as
\begin{equation}
U_{pqr}(\gamma) = \exp\Big(\ii\oint_{\gamma} pA_1+qA_2+rA_3\Big),~p,q,r=0,1
\end{equation}
which are gauge invariant under gauge transformations Eq.~\eqref{eq: gauge transf 2+1}. Using the form of the Lagrangian \eqnref{eq: lag 2+1}, the compactness condition on $b_i$ in \eqnref{eq: comp transf 2+1} breaks $A_i$ down to $\mathbb{Z}_2$, so the operator $U_{pqr}$ only depends on the values of $p,q,r$ modulo $2$.

Flux insertion operators are more complicated, since a single term $\exp(\ii\oint_{\gamma} b_i),~i=1,2,3$ is not gauge invariant under transformation \eqnref{eq: gauge transf 2+1}. One can construct the flux insertion operators by introducing auxiliary fields $\phi_i$ and $\lambda_i$ living on the loop $\gamma$, Indeed, we find that the following operator defined via a path integral is gauge invariant
\begin{equation}\label{eq.100}
\begin{split}
V_{100}(\gamma)=&\frac{1}{\mathcal{N}} \int D[\phi_2]D[\phi_3]D[\lambda_2]D[\lambda_3]\\	\exp\Bigg(\ii\oint_{\gamma}
& b_1 + \sum_{ij=2}^{3} \frac{\epsilon^{1ij}}{\pi} (\frac{1}{2} \phi_i d \phi_j + (d\phi_i - A_i) \lambda_j )\Bigg)\punc{,}
\end{split}
\end{equation}
where $i$ and $j$ are actually summed over $\{2,3\}$ because of $\epsilon^{1ij}$; $\mathcal{N}$ is a normalization factor which we determined later in \secref{subsec: fusion rules 2+1} by insisting on a consistent set of fusion rules for the flux insertion operators. The operator of Eq.~\eqref{eq.100} is gauge invariant under gauge transformation Eq.~\eqref{eq: gauge transf 2+1} with additional transformations, $\phi_i \mapsto \phi_i + \alpha_i$ and $\lambda_i \mapsto \lambda_i + \alpha_i$.

The auxiliary fields in Eq.~\eqref{eq.100}, $\phi_2$, $\phi_3$, $\lambda_2$ and $\lambda_3$ can be integrated out exactly -- the details of the calculation can be found in the App.~\ref{app. path integral}. The result is conveniently expressed as
\begin{equation}\label{eq.100 close form}
V_{100}=2 \exp\Big(\ii\oint_{\gamma} b_1 + \sum_{i,j=1}^{3} \frac{\epsilon^{1ij}}{2\pi} \w_i d \w_j\Big) \delta(\bar{\w}_2|_{\gamma})\delta(\bar{\w}_3|_{\gamma}),
\end{equation}
where $\w_i$ is the holonomy function for $A_i$ which is defined explicitly on the loop $\gamma$ as
\be\label{eq:wi}
\w_i (x):= \int_{\gamma, x_0}^x A_i\punc{,}
\ee
while $\bar{\w}_i:=\oint_{\gamma} A_i$, $i=1,2,3$. The choice of  origin of integration $x_0$ is arbitrary. The $\delta$ functions appearing in \eqnref{eq.100 close form}  project onto configurations for which the $A_2,A_3$ fluxes threading $\gamma$ are zero. They are not the usual  $\delta$ functions encountered in the continuum -- rather they are defined to be a projector to the trivial holonomy state: $\delta(\bar{\w}_i |_\gamma):=\frac{1}{2} (1+ \exp(\ii \bar{\w}_i |_\gamma))$. By trivial holonomy, we mean $ \bar{\w}_i=2\pi n$ for any $n$, where $n$ is an integer. We will come back to the overall factor of $2$ in \eqnref{eq.100 close form} when fusion rules are discussed in Sec. \ref{subsec: fusion rules 2+1}. Note that the expression $\oint_{\gamma} \w_i d \w_j$ resulting from integrating out the scalar fields is not local in terms of the gauge fields $A_i$. The other flux insertion operators have similar expressions

\begin{widetext}
\begin{equation}\label{eq.010}
\begin{split}
V_{010}(\gamma)=&\frac{1}{\mathcal{N}} \int D[\phi_1]D[\phi_3]D[\lambda_1]D[\lambda_3]	\exp\Big(\ii\oint_{\gamma} b_2 + \sum_{i,j=1}^{3} \frac{\epsilon^{2ij}}{\pi} (\frac{1}{2} \phi_i d \phi_j + (d\phi_i - A_i) \lambda_j )\Big)	\\
=& 2 \exp\Big(\ii\oint_{\gamma} b_2 + \sum_{i,j=1}^{3} \frac{\epsilon^{2ij}}{2\pi} \w_i d \w_j\Big) \delta(\bar{\w}_1|_{\gamma})\delta(\bar{\w}_3|_{\gamma})	\;,
\end{split}
\end{equation}
\begin{equation}\label{eq.001}
\begin{split}
V_{001}(\gamma)=&\frac{1}{\mathcal{N}} \int D[\phi_1]D[\phi_2]D[\lambda_1]D[\lambda_2]	\exp\Big(\ii\oint_{\gamma} b_3 + \sum_{i,j=1}^{3} \frac{\epsilon^{3ij}}{\pi} (\frac{1}{2} \phi_i d \phi_j + (d\phi_i - A_i) \lambda_j )\Big)	\\
=& 2 \exp\Big(\ii\oint_{\gamma} b_3 + \sum_{i,j=1}^{3} \frac{\epsilon^{3ij}}{2\pi} \w_i d \w_j\Big) \delta(\bar{\w}_1|_{\gamma})\delta(\bar{\w}_2|_{\gamma}).
\end{split}
\end{equation}
\end{widetext}

Before moving on to the remaining flux insertion operators, let us further motivate the path integral form of the operators $V_{100}$, Eq.~\eqref{eq.100}, and similarly for $V_{010}$ Eq.~\eqref{eq.010} and $V_{001}$ Eq.~\eqref{eq.001}. Gauge invariance strongly constrains the forms of these operators. If we write down an operator of the form
\begin{equation}
V_{100} = \exp(\ii\oint_\gamma b_1) \mathfrak{g}(A),
\end{equation}
and insist on gauge invariance, we find that the functional $\mathfrak{g}$ is necessarily a non-local functional of $A$ -- it must have something like the $A_i$ dependence of \eqnref{eq.100 close form}, involving constraints $\bar{\w}_2=\bar{\w}_3=0$, and phase terms like $\int_\gamma \w_i d \w_j$. In order to realize the operator Eq.~\eqref{eq.100 close form} in a local form, one possible solution is to introduce auxilliary fields into the path integral living on $\gamma$ which once integrated out, realize \eqnref{eq.100 close form}. This is the approach which led to \eqnref{eq.100}.

While the introduction of these auxiliary fields may seem ad hoc, there is a neat underlying physical interpretation for this procedure. To understand this interpretation, we briefly return to  the quantum double theory approach of Ref.~\onlinecite{propitius1995topological}.  Within that algebraic framework, the flux quasi-particles  for the $\mathbb{Z}_2\times \mathbb{Z}_2 \times \mathbb{Z}_2$ theory considered here carry a projective representation. In other words, the fluxes carry an internal degree of freedom which transforms projectively under the gauge group.  The flux insertion operators we consider insert precisely such fluxes, so should also carry some such internal degree of freedom. And indeed they do: One way of interpreting the $\phi,\lambda$ fields is that they are matter fields which on net transform projectively under the gauge group.

To further substantiate this idea, note that in the study of SPT phases, the boundary of a (1+1)D SPT bulk transforms projectively under the bulk symmetry\cite{pollmann2012symmetry,chen2011classification,fidkowski2011topological}. This statement, curiously enough, is helpful in interpreting our line operators. Suppose we have an abstract form of $V_{100}$ as follows:
\begin{equation}
V_{100} = \int D[\phi] D[\lambda] \ldots \exp(\ii\oint_\gamma b_1 + f(A,\phi,\lambda,\ldots)),
\end{equation}
where $f$ is a function of $A$ and auxiliary fields such as $\phi$, $\lambda$ etc. And we try to calculate the expectation value of $V_{100}$. We need:
\begin{widetext}
\begin{equation}\label{eq. DecoratedDomainWall}
\begin{split}
\langle V_{100} \rangle =& \int D[b_i] D[A_i] D[\phi] D[\lambda] \exp(\ii \int \lag + \ii \oint_\gamma (b_1 + f(A,\phi,\lambda,\ldots )))	\\
=& \int D[A_i] D[\phi] D[\lambda] \exp(\ii \int \frac{1}{\pi^2} \bar{A}_1\bar{A}_2\bar{A}_3 + \ii \oint_\gamma f(\bar{A},\phi,\lambda,\ldots ))	\\
=& \int D[A_i] D[\phi] D[\lambda] \exp(\ii \int_{[\bar{A}_1]} \frac{1}{\pi} \bar{A}_2\bar{A}_3	+ \ii \oint_\gamma f(\bar{A},\phi,\lambda,\ldots )).
\end{split}
\end{equation}
\end{widetext}
We have omitted all the wedges ``$\wedge$" in the above, and will continue this convention in the following texts if without misunderstanding.
The second equality comes from integrating out all $b_i$ fields. In this case, $A_2$ and $A_3$ will be flat and thus be exact on a simple spacetime manifold, while $A_1$ will not. And these fields after integrating out $b_i$ fields are denoted as $\bar{A}_1$, $\bar{A}_2$ and $\bar{A}_3$. Note that the integral $\int \bar{A}_1 \bar{A}_2\bar{A}_3$ can be written as the integral over $A_1$ flux sheet $[\bar{A}_1]$ (see Ref.~\onlinecite{Ye:2015eba} for a similar discussion) whose boundary is $\partial[\bar{A}_1]=\gamma$:
\begin{eqnarray}
\int \bar{A}_1\bar{A}_2\bar{A}_3 = \int_{[\bar{A}_1]} \pi\bar{A}_2\bar{A}_3 \;.
\end{eqnarray}
where $\pi$ comes from the normalization. We still need the rest of the terms in the second equality of Eq.~\eqref{eq. DecoratedDomainWall} to be gauge invariant. Then the gauge anomalies of two integrals $\int_{[\bar{A}_1]} \frac{1}{\pi} \bar{A}_2\bar{A}_3$ and $\oint_\gamma f(\bar{A},\phi,\lambda,\ldots)$ need to cancel each other. Notice that $\int_{[\bar{A}_1]} \bar{A}_2\bar{A}_3$ is just the SPT Lagrangian on the manifold $[\bar{A}_1]$ with symmetry $\mathbb{Z}_2^{\otimes2}$, when $\bar{A}_2$ and $\bar{A}_3$ are both exact. Hence $f$ should look like the boundary action of an SPT, insofar as it should transform to compensate for the gauge anomaly from the bulk action. Indeed, the particular $f$ chosen in \eqnref{eq.100} looks very much like the boundary action of the SPT in Ref.~\onlinecite{kapustin2014coupling}.


We can similarly write down the direct generalizations of flux insertion operators from Eq.~\eqref{eq.100}, which insert two types of fluxes and three types of flues.
\begin{widetext}
\begin{equation}\label{eq.110}
\begin{split}
V_{110}=&\fr{1}{\mathcal{N}'}\int D[\phi_i]D[\lambda_i] \exp\Big(\ii\oint_{\gamma} b_1 + b_2 + \sum_{i=1,2}\sum_{j,k=1}^{3}\frac{\epsilon^{ijk}}{\pi} (\frac{1}{2}\phi_j d \phi_k + (d\phi_j-A_j)\lambda_k)\Big)	\\
=& 2 \exp\Big(\ii\oint_{\gamma} b_1 + b_2 + \sum_{i=1,2}\sum_{j,k=1}^{3} \frac{\epsilon^{ijk}}{2\pi}\w_j
d\w_k\Big) \delta(\bar{\w}_1|_{\gamma}-\bar{\w}_2|_{\gamma})\delta(\bar{\w}_3|_{\gamma}),
\end{split}
\end{equation}
\begin{equation}\label{eq.111}
\begin{split}
V_{111}=&\fr{1}{\mathcal{N}''}\int D[\phi_i]D[\lambda_i] \exp\Big(\ii\oint_{\gamma} b_1 + b_2 + b_3 + \sum_{i,j,k=1,2,3}\frac{\epsilon^{ijk}}{\pi} (\frac{1}{2}\phi_j d \phi_k + (d\phi_j-A_j)\lambda_k)\Big)	\\
=& 2 \exp\Big(\ii\oint_{\gamma} b_1 + b_2 + b_3 + \sum_{i,j,k=1,2,3} \frac{\epsilon^{ijk}}{2\pi}\w_j
d\w_k\Big) \delta(\bar{\w}_1|_{\gamma}-\bar{\w}_2|_{\gamma})\delta(\bar{\w}_2|_{\gamma}-\bar{\w}_3|_{\gamma}).
\end{split}
\end{equation}
\end{widetext}
The notations are the same as in Eq.~\eqref{eq.100} and \eqref{eq.100 close form}. The second equalities of both the above equations follow by integrating out all $\phi_i$ and $\lambda_i$ fields. Start by  integrateing out $\lambda_1$. Then we have a constraint $d\phi_2 - d\phi_3 = A_2 - A_3$, the solution of which can be written as $\phi_2=\phi_3 + \w_2 - \w_3 + C_2$ where $C_2$ is a constant. Similarly integrating out $\lambda_2$ yields $\phi_1=\phi_3 + \w_1 - \w_3 + C_1$. Lastly, the constraint obtained by integrating out $\lambda_3$ is automatically satisfied. Plugging these two solutions back in produces the second equality, where $C_2$ and $C_3$ have been shifted away. Notice that the solutions for the two constraints exists with the condition that $\bar{\w}_1=\bar{\w}_2=\bar{\w}_3$, which is actually a similar phenomenon in the cases of $V_{110}$ and $V_{111}$.

The operators $V_{100}$, $V_{110}$ and $V_{111}$ share formal similarities as we have seen from their closed form. However, they  differ from $V_{100}$, $V_{010}$ and $V_{001}$ in that the projector $\delta$ function changes. In the operator $V_{110}$, Eq.~\eqref{eq.110}, we need the projector that forces $\bar{\w}_1=\bar{\w}_2$, while $\bar{\w}_3$ is forced to be trivial. Similarly for $V_{101}$ and $V_{011}$. In the operator $V_{111}$, Eq.~\eqref{eq.111}, $\bar{\w}_1$, $\bar{\w}_2$ and $\bar{\w}_3$ are forced to be the same by the $\delta$ function. These $\delta$ functions  will be essential when we compute the correlation functions in Sec.~\ref{subsec: correlation 2+1}.

We have listed all possible Wilson operators and flux insertion operators, using gauge invariance and locality as our principle constraints. Our ansatz is inspired by considering anomaly inflow in lower dimensions. We have found  8 types of Wilson operators $U_{pqr}$, $(p,q,r=0,1)$, and 8 types of flux insertion operators $V_{pqr}$, $(p,q,r=0,1)$. Therefore there should be 64 types line operators including all the composites of Wilson and flux insertion operators. However, in the next section, Sec.~\ref{subsec: 22 line operators}, we will show that many operators are identified due to the $\delta$ function, and there are only 22 distinguishable line operators in total. This agrees with the  quantum double calculation in Ref.~\onlinecite{propitius1995topological}.

\subsection{22 Distinguishable Line Operators}
\label{subsec: 22 line operators}
In this section, we will show that in the type III twisted $\mbb{Z}_2^{\otimes 3}$ field theory Eq.~\eqref{eq: lag 2+1}, there are only 22 distinguishable line operators rather than naively 64 line operators. To show this, we argue that some operators always have the same correlation functions. Hence, many of the naive $64$ operators should be identified since they have identical correlation functions with all other operators. The essential point is that $V_{pqr}$ are always associated with certain constraints ($\delta$ functions) on the gauge fields $A_i$. See the $\delta$ functions in Eq.~\eqref{eq.100 close form}, \eqref{eq.110} and \eqref{eq.111} resulting from integrating out the matter fields $\lambda_i,\phi_i$. As a result, the insertion of a flux insertion operator along loop $\mathcal{C}$ fixes certain combinations of holonomies of the gauge fields along the same loop $\mathcal{C}$. The flux insertion operator then has trivial fusion rules with Wilson lines corresponding to the mentioned holonomies, simply because the flux insertion operator fixes the values of the Wilson lines. So fusing the flux insertion line with certain Wilson lines is precisely the same as inserting just the flux insertion line.

Let us argue more concretely with an example. We have already listed 8 pure Wilson operators $U_{pqr}$ which insert charges,  and 8 pure flux insertion operators $V_{pqr}$ which insert fluxes. We now consider composites of the two kinds of operator. First consider $V_{100}$ along loop $\gamma$ and fuse it with Wilson operator $U_{p q r}$. One can compute the correlation function of the composite operator with arbitrary operator $\langle \mathcal{O}V_{100}\times U_{p q r}\rangle$, and measure the effect of the additional Wilson operator. We assume that the support of the operator $\mathcal{O}$ excludes $\gamma$.

Multiplying $V_{100}$ by $U_{010}(\gamma)$ or $U_{001}(\gamma)$ or their combination will not change the correlation function, because $A_2$ and $A_3$ fields have trivial holonomy along $\gamma$ -- this follows from the $\delta$ function constraint in \eqnref{eq.100 close form}, which directly implies $\exp(\ii\oint_{\gamma}A_2)$ and $\exp(\ii\oint_{\gamma}A_3)$ equal 1. Then $U_{010}=1$ and $U_{001}=1$ within the correlation functions $\langle V_{100}(\gamma) U_{010}(\gamma) \mathcal{O} \rangle$ and $\langle V_{100}(\gamma)U_{001}(\gamma)\mathcal{O} \rangle$ for any $\mathcal{O}$.

On the other hand, there is no constraint on $\bar{\w}_1$ in $V_{100}$, so the holonomy of $A_1$ around $\gamma$ is unconstrained, and indeed we can (and will) construct operators $\mathcal{O}$ such that $V_{100} \times U_{100}\mathcal{O}\neq V_{100} \mathcal{O}$ within a correlation function. To summarize, we find that
\begin{equation}
\begin{split}
\langle V_{100}(\gamma)\mathcal{O}\rangle
=&\langle V_{100}(\gamma)U_{010}(\gamma)\mathcal{O}\rangle	\\
=&\langle V_{100}(\gamma)U_{001}(\gamma)\mathcal{O}\rangle,	\\
=&\langle V_{100}(\gamma)U_{010}(\gamma)U_{001}(\gamma)\mathcal{O}\rangle	\\
\langle V_{100}U_{100}(\gamma)\mathcal{O}\rangle
=&\langle V_{100}U_{100}(\gamma)U_{010}(\gamma)\mathcal{O}\rangle	\\
=&\langle V_{100}U_{100}(\gamma)U_{001}(\gamma)\mathcal{O}\rangle	\\
=&\langle V_{100}U_{100}(\gamma)U_{010}(\gamma)U_{001}(\gamma)\mathcal{O}\rangle\punc{.}
\end{split}
\end{equation}

Therefore, all of the distinguishable operators associated with $V_{100}$  are divided into two equivalence classes --that is, they have the same correlation functions as one of $V_{100}(\gamma)$, $V_{100}(\gamma)U_{100}(\gamma)$.  We adopt the quantum double notation by denoting the two classes of $V_{100}(\gamma)$ and $V_{100}(\gamma)U_{100}(\gamma)$  as $(100,\alpha_{\pm}^1)$ respectively, where $100$ represents $V_{100}$ and the plus sign corresponds to $V_{100}(\gamma)$ while the minus sign corresponds to $V_{100}(\gamma)U_{100}(\gamma)$.

Similar arguments can also be applied for $V_{010}$ and $V_{001}$, where the operators are denoted similarly by $(010,\alpha_{\pm}^2)$, $(001,\alpha_{\pm}^3)$.

Using the same ideas, we consider fusing $V_{110}(\gamma)$ with various Wilson lines. Once again the $\delta$ function constraints arising from integrating out the $\phi,\lambda$ matter fields  Eq.~\eqref{eq.110} are useful. In this case, the constraints imply that $A_1$ and $A_2$ share the same holonomy along $\gamma$, while $A_3$ has no holonomy along $\gamma$. As a result $V_{110}U_{100}$ always gives the same correlation functions as $V_{110}U_{010}$ does, and $V_{110}U_{110}$ and $V_{110}U_{001}$ give the same correlation functions as $V_{110}$. Therefore, we there are two equivalence classes of $V_{110}$ operator with representatives (for example) $V_{110}$ and $V_{110}U_{100}$. We denote them by $(110, \beta^3_{\pm})$ respectively. The same line of reasoning  also applies to $V_{011}$ and $V_{101}$. We denote the operators by a similar notation, $(011, \beta^1_{\pm})$ and $(101, \beta^2_{\pm})$.

Finally, let us consider the possible fusions of $V_{111}(\gamma)$ with Wilson lines. Using the constraint in \eqnref{eq.111},  $A_1$, $A_2$ and $A_3$ must share the same holonomy along $\gamma$. Therefore, we find that  $U_{100}$, $U_{010}$, $U_{001}$ and $U_{111}$ are equivalent along $\gamma$. Moreover, $U_{110}$, $U_{011}$ and $U_{101}$ are all equal to $1$ and do not contribute any phases to the correlation functions. Therefore, once again there are two equivalence classes of line operators which we denote $(111,\gamma_{\pm})$, where ``$+$" sign corresponds to $V_{111}$ itself, or its decorations by $U_{110}$, $U_{011}$ and $U_{101}$, and ``$-$" sign corresponds to the equivalence class $V_{111}U_{100}$, $V_{111}U_{010}$, $V_{111}U_{001}$ and $V_{111}U_{111}$.

In summary, we have $22$ distinguishable operators in total: $U_{pqr}~(p,q,r=1,2,3)$, $(100,\alpha^1_{\pm})$, $(010,\alpha^2_{\pm})$, $(001,\alpha^3_{\pm})$, $(011, \beta^1_{\pm})$, $(101, \beta^2_{\pm})$, $(110, \beta^3_{\pm})$, and $(111,\gamma_{\pm})$. The same result also arises from quantum double calculation with type III cocycles. C.f. Ref.~\onlinecite{propitius1995topological}. We have therefore established a 1-to-1 map between field theoretical operators and the projective representations in quantum double models.

\subsection{Fusion Rules of Line Operators}
\label{subsec: fusion rules 2+1}
Having identified the various possible gauge invariant line operators, we calculate their fusion rules.  This allows us to motivate the normalizations used in defining the line operators, e.g., the factor of two appearing in Eq.~\eqref{eq.100 close form}.

In quantum field theory, the fusion of two line operators is defined via the process of dragging two lines operators close to each other. The outcome of the product of two line operators can be decomposed as a sum of a set of line operators. If the fusion outcome can only contain one operator, we will call such theories and operators, ``abelian" theories and ``abelian particles" respectively . Similarly, we will call them ``non-abelian" theories and ``non-abelian" particles if there exists more than one fusion outcome.

To begin with, we can calculate fusion rules of Wilson operators quite straightforwardly
\begin{equation}\label{eq.U fuse U}
U_{pqr}(\gamma) \times U_{xyz}(\gamma) = U_{(q+x)(q+y)(r+z)}(\gamma)\punc{,}
\end{equation}
where the sums are defined modulo 2. The fusion rules, Eq.~\eqref{eq.U fuse U}, also demonstrate that all Wilson operators are abelian.

Next we address the flux insertion operators. Henceforth, for simplicity, we  adopt the closed form of flux insertion operators $V_{pqr}$ written in terms of the holonomy functions $\w_i$'s. For example Eq.~\eqref{eq.100 close form}. To begin, let us fuse the same two flux insertion operators $(100,\alpha^1_+)$
\begin{equation}
(100,\alpha^1_+)\times(100,\alpha^1_+) \equiv V_{100}\times V_{100}\punc{,}
\end{equation}
where we have used the fact established in the last section that $V_{100}$ is a representative in the class of $(100,\alpha^1_+)$. By definition of $V_{100}$ in Eq.~\eqref{eq.100 close form}, we have
\begin{equation}\label{eq.100 fuse 100}
\begin{split}
& V_{100}\times V_{100}	\\
=& 4\exp(\ii\oint_{\gamma} 2b_1+\fr{2}{\pi}\w_2d\w_3)) (\delta(\bar{\w}_2|_{\gamma}))^2(\delta(\bar{\w}_3|_{\gamma}))^2	\\
=& 4\delta(\bar{\w}_2|_{\gamma})^2\delta(\bar{\w}_3|_{\gamma})^2	\\
=& 4\delta(\bar{\w}_2|_{\gamma})\delta(\bar{\w}_3|_{\gamma})	\\
=& (1+\exp(\ii\bar{\w}_2))(1+\exp(\ii\bar{\w}_3))	\\
=& 1+\exp(\ii\bar{\w}_2)+\exp(\ii\bar{\w}_3)+\exp(\ii\bar{\w}_2)\exp(\ii\bar{\w}_3)	\\
=& U_{000}+U_{010}+U_{001}+U_{011}.
\end{split}
\end{equation}
This rather bizarre looking calculation requires some explanations. The first equality just follows from definition of $V_{100}$, and the $\delta$ function is actually a projector that projects into zero flux state (more explicitly, $\delta(\bar{\w}_i|_{\gamma}):=\frac{1}{2}(1+\exp(\ii\bar{\w}_i)),~(i=1,2,3)$ as noted below \eqnref{eq:wi}); the second equality follows from the fact that all variables are $\mbb{Z}_2$ variables valued in $\{0,\pi\}$, then the exponential is actually trivial because it is always $2\pi$ \footnote{ It may seem strange to discuss a $\mbb{Z}_2$ in the continuum. In fact the expressions here are short-hand for a more careful (but more cumbersome) calculation on the lattice, where there is certainly no obstacle to considering $\mathbb{Z}_2$ and more generally discrete valued fields.}.  The third equality follows from the fact that the $\delta$ function satisfies $(\delta(\bar{\w}_i))^2=\delta(\bar{\w}_i)$; the fourth equality just expresses the $\delta$ functions explicitly as $\delta(\bar{\w}_i|_{\gamma})=\frac{1}{2}(1+\exp(\ii\bar{\w}_i))$.

Using quantum double notation, Eq.~\eqref{eq.100 fuse 100} is expressed as
\begin{equation}\label{eq:integerfusion}
(100,\alpha^1_+)\times(100,\alpha^1_+)=1 + U_{010} + U_{001} + U_{011}.
\end{equation}
As promised in \secref{subsec: line operators 2+1}, we need to motivate the normalization factors for the flux insertion operators. Indeed, the fact insisting that fusion rules like \eqnref{eq:integerfusion} involve positive integer combinations of line operators fixes the overall normalization factors (e.g., the $2$ factor in Eq.~\eqref{eq.100}).

As another example, consider  fusion rule
\begin{equation}
(010,\alpha^2_+)\times(001,\alpha^3_+)\equiv V_{010}\times V_{001}\punc{.}
\end{equation}
To see how this comes about, we use our explicit expressions for the line operators (Eq.~\eqref{eq.010} and Eq.~\eqref{eq.001})
\begin{equation}
\begin{split}\label{eq:010x001}
V_{010}\times V_{001}=&4\exp\Big[\ii\oint_{\gamma} b_2+b_3+\fr{1}{\pi}(-\w_1 d \w_3 + \w_1 d \w_2)\Big]	\\
& \delta(\bar{\w}_1|_{\gamma})\delta(\bar{\w}_2|_{\gamma})\delta(\bar{\w}_3|_{\gamma})\punc{.}
\end{split}
\end{equation}
The right hand side of this equation can be manipulated into the form
\begin{equation}
V_{010}\times V_{001}=V_{011}+V_{011}U_{010}=V_{011}+V_{011}U_{001}\punc{.}
\end{equation}
To see why, in \eqnref{eq:010x001} rewrite $\delta(\bar{\w}_2 |_\gamma) \delta(\bar{\w}_3 |_\gamma) = \delta(\bar{\w}_2 |_\gamma - \bar{\w}_3 |_\gamma) \delta(\bar{\w}_3 |_\gamma)$ and expand $\delta(\bar{\w}_3|_{\gamma})=\frac{1}{2} (1+ \exp(\ii \bar{\w}_3 |_\gamma))$. Then compare the result with the definition of $V_{011}$ from Eq.~\eqref{eq.110}
\begin{equation}
\begin{split}
V_{011}=& 2\exp\Big[i\oint_{\gamma} b_2+b_3 +\fr{1}{\pi}(-\w_1 d \w_3 + \w_1 d \w_2)\Big]	\\
&\delta(\bar{\w}_1|_{\gamma})\delta(\bar{\w}_2|_{\gamma}-\bar{\w}_3|_{\gamma})\punc{.}
\end{split}
\end{equation}
In terms of the quantum double notation we have shown that
\begin{equation}
(010,\alpha^2_+) \times (001,\alpha^3_+) = (011,\beta^1_+) + (011,\beta^1_-)\punc{.}
\end{equation}

We can readily find the quantum dimensions of all of the operators above. The Wilson lines all have quantum dimension $1$. This follows most readily from the fact that $U_{pqr}\times U_{pqr} = 1$ in \eqnref{eq.U fuse U}, along with some  general constraints on the structure of fusion algebras \footnote{Use Eq. 14 of Ref.~\onlinecite{bernevig2015topological}.}. Moreover, the fusion $(100, \alpha^+)\times (100, \alpha^+)$ in \eqnref{eq:integerfusion} gives a sum of four Wilson lines. Again using Ref.~\onlinecite{bernevig2015topological}, this implies that the quantum dimension of $(100, \alpha^+)$ is 2. Similarly, for other flux insertion operators.  The overall factor $2$ in the definitions of the flux insertion lines operators, for example Eq.~\eqref{eq.100 close form}, is actually the quantum dimension for the operators.

In summary, we have worked out several examples of  fusion rules and quantum dimensions of the line operators using our field theoretic formalism. We have demonstrated that the type III twisted $\mathbb{Z}_2^{\otimes3}$ theory is a non-abelian topological phase, even though it is a topological field theory involving only abelian gauge fields.

\subsection{Correlation Functions of Line Operators}
\label{subsec: correlation 2+1}
\begin{figure}[b]
\centering
\includegraphics[width=0.4\textwidth]{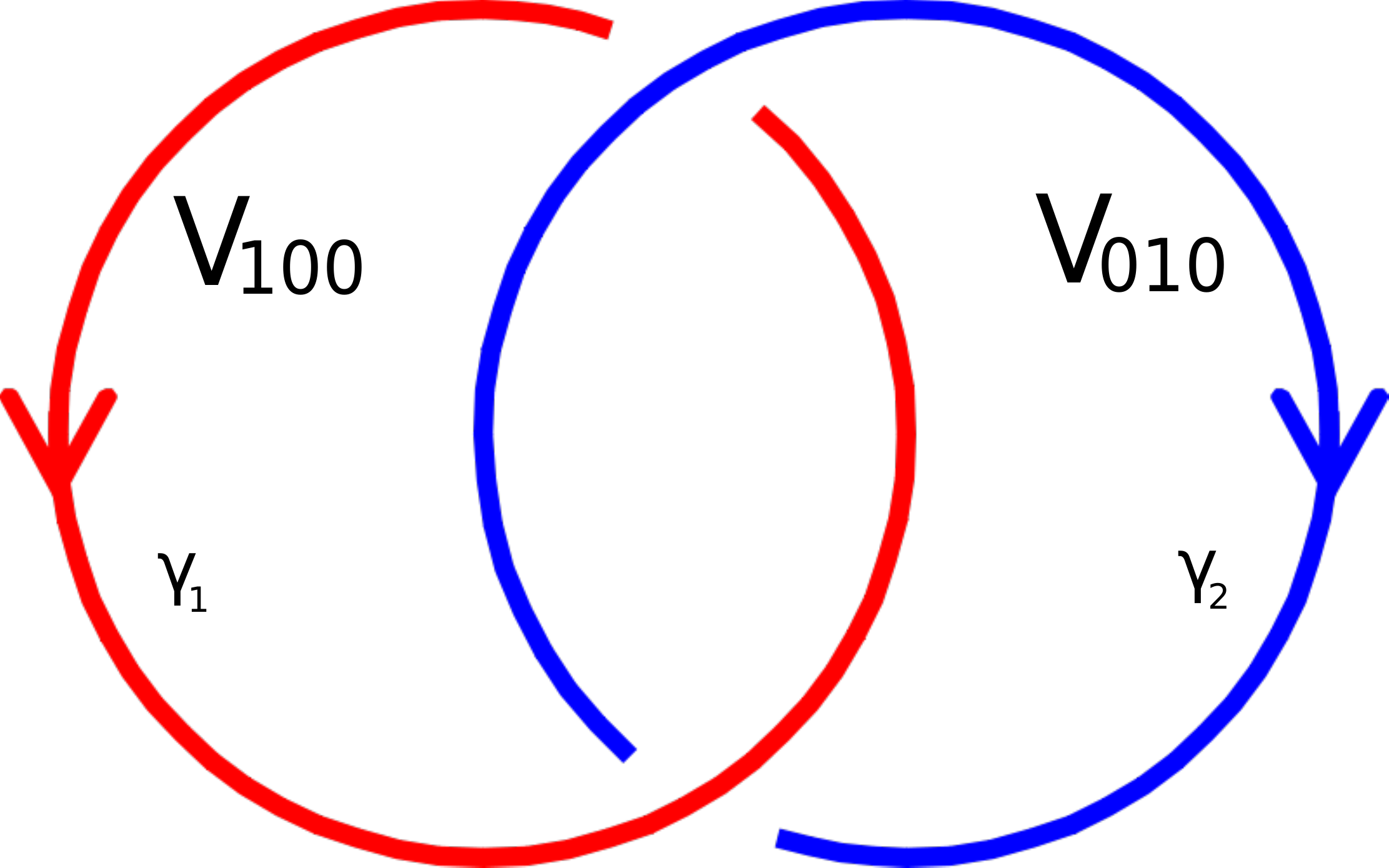}
\caption{An illustration of a linking correlation of $V_{100}(\gamma_1)$ and $V_{010}(\gamma_2)$.}
\label{fig. Hopf}
\end{figure}

In this section, we calculate correlation functions for line operators that link one another. Typically, we will consider two line operators forming a Hopf link in (2+1)D, Fig.~\ref{fig. Hopf}. If we have a link of two t' Hooft operators corresponding to gauge fluxes $\phi_1,\phi_{2}$, then the holonomy along the first loop is $\phi_1$ while that along the second loop is $\phi_{2}$. As we have seen before, the flux insertion operators are associated with constraints on the holonomies of $A_i$ along the loop. As a result, we will see that for many of the possible links the holonomies are not compatible with the constraints, so the expectation value for the link is simply zero.

To demonstrate this point, examine a link of $V_{100}(\gamma_1)$ and $V_{010}(\gamma_2)$: $V_{010}(\gamma_2)$ will insert a holonomy $\pi$ of $A_2$ along $\gamma_1$. See Fig~\ref{fig. Hopf}. However, we know from \eqnref{eq.100} that $V_{100}(\gamma_1)$ is associated with two constraints $\oint_{\gamma_1} A_2=\oint_{\gamma_1} A_3=0$. The mismatch between the holonomy and the constraint leads to a zero expectation value. For explicit path integral calculation details for the linking correlation of $V_{100}(\gamma_1)$ and $V_{010}(\gamma_2)$, please refer to App.~\ref{app. path integral}.

The mismatching of the constraints and flux insertion mean that most of the Hopf links we consider disappear. Here in the main text, we present only one subtle calculation, the corellator $\langle V_{111}(\gamma_1)V_{111}(\gamma_2)\rangle $ where $\gamma_1$ and $\gamma_2$ form a link in (2+1)D.
\begin{widetext}
\begin{equation}\label{eq.111correlation}
\begin{split}
&\langle V_{111}(\gamma_1) V_{111}(\gamma_2) \rangle	\\
=& 4 \int Db_i DA_i \exp(\ii S_0) \exp(\ii\oint_{\gamma_1} (b_1+b_2+b_3)+\fr{\ii}{2\pi}\oint_{\gamma_1}\sum_{ijk}\epsilon^{ijk} \w_i d \w_j)
\delta(\bar{\w}_1|_{\gamma_1}-\bar{\w}_2|_{\gamma_1})\delta(\bar{\w}_2|_{\gamma_1}-\bar{\w}_3|_{\gamma_1})	\\
& \exp(\ii\oint_{\gamma_2} (b_1+b_2+b_3)+\fr{\ii}{2\pi}\oint_{\gamma_2}\sum_{ijk}\epsilon^{ijk} \w_i d w_j)
\delta(\bar{\w}_1|_{\gamma_2}-\bar{\w}_2|_{\gamma_2})\delta(\bar{\w}_2|_{\gamma_2}-\bar{\w}_3|_{\gamma_2})	\\
=& 4 \exp(\ii\int \frac{1}{\pi^2} \tilde{A}_1\tilde{A}_2\tilde{A}_3) \exp(\fr{\ii}{2\pi}\oint_{\gamma_1}\sum_{ijk}\epsilon^{ijk} \tilde{\w}_i d \tilde{\w}_j) \exp(\fr{\ii}{2\pi}\oint_{\gamma_2}\sum_{ijk}\epsilon^{ijk} \tilde{\w}_i d \tilde{\w}_j)	\\
=& 4 \exp(\fr{\ii}{2\pi}\oint_{\gamma_1}\sum_{ijk}\epsilon^{ijk} \tilde{\w}_i d \tilde{\w}_j) \exp(\fr{\ii}{2\pi}\oint_{\gamma_2}\sum_{ijk}\epsilon^{ijk} \tilde{\w}_i d \tilde{\w}_j).
\end{split}
\end{equation}
\end{widetext}
The first equality follows from our definitions of the line operators. The second equality is obtained by integrating out all $b_i$ fields. This yields the constraints
\begin{equation}
\frac{1}{\pi}dA_i=\star (j(\gamma_1)+j(\gamma_2)),
\end{equation}
where $j(\gamma_1)$ and $j(\gamma_2)$ are the unit vectors tangential to $\gamma_1$ and $\gamma_2$ respectively. Here $\star$ is the Hodge dual. For positions away from the support of the loops $\gamma_{1,2}$ we have $dA_i=0$. We denote by $\tilde{A}_i$ a particular reference solution to the constraints on $A_i$. One thing we certainly know about the $\tilde{A}_i$ is that they have nontrivial holonomies along $\gamma_1$ and $\gamma_2$, i.e.
\begin{equation}
\oint_{\gamma_1} \tilde{A}_i=\oint_{\gamma_2} \tilde{A}_i=\pi \mod 2\pi ,~\forall i=1,2,3.
\end{equation}
Therefore, the $\delta$ functions in the first equality in \eqnref{eq.111correlation} is automatically satisfied. Moreover, since all $A_i$ obey the same equations of motion set by integrating out $b_i$ fields, their solutions $\tilde{A}_i$ are the same up to gauge transformations. Therefore, we are free to choose a gauge for which the $\tilde{A}_i$ are identical, so that the integral of $\frac{1}{\pi^2} \tilde{A}_1\tilde{A}_2\tilde{A}_3$ vanishes at least in this quasi-continuum setting. All we need to do now is to evaluate the last two line integrals  in the last line of Eq.~\eqref{eq.111correlation}.

In the last line, $\tilde{\w}_i$'s are the holonomy functions for each field $A_i$. As we have chosen a gauge for which the $\tilde{A}_i$ are identical, the corresponding holonomy functions are identical.  As a result, on each loop the  $\tilde{\w}_i$ are identical multi-valued staircase functions which sharply step up by $\pi$ modulo $2\pi$ upon moving around the loop once.

In order to evaluate these line integrals, we need to think more carefully about the regularization of the field theory. To this end, we consider using discrete derivatives on a lattice of form
\begin{equation}
\begin{split}
d\tilde{\w}_i(r) =& \tilde{\w}_i(r+1) - \tilde{\w}_i(r)	\\
\bar{d}\tilde{\w}_i(r) =& \tilde{\w}_i(r) - \tilde{\w}_i(r-1),	
\end{split}
\end{equation}
where $\bar{d}$ is the adjoint operator to $d$ on the lattice, $r$ is the position on the lattice.  In terms of these operators, the line integrals over $\gamma_1$ and $\gamma_2$ are regularized as
\begin{equation}
\begin{split}
& \oint_{\gamma_1/\gamma_2} \tilde{\w}_i d \tilde{\w}_j - \tilde{\w}_j \bar{d} \tilde{\w}_i	\\
=&\sum_{r=-L/2}^{L/2} \lbrack \tilde{\w}_i(r)(\tilde{\w}_j(r+1)-\tilde{\w}_j(r)) 	\\
&-  \tilde{\w}_j(r)(\tilde{\w}_i(r)-\tilde{\w}_i(r-1)) \rbrack	\\
=& \tilde{\w}_i(-1)\pi - \tilde{\w}_j(0)\pi	\\
=& -\pi^2.
\end{split}
\end{equation}
More details of the particular choice of derivatives can be found in the Sec.~\ref{sec. ZN examples} when we verify that the operators for type III twisted $\mbb{Z}_N^{\otimes3}$ theories are gauge invariant on lattice. Hence we have
\begin{equation}
\begin{split}
\sum_{ijk} \epsilon^{ijk} \tilde{\w}_i d\tilde{\w}_j \equiv & (\tilde{\w}_2 d\tilde{\w}_3 - \tilde{\w}_3 \bar{d}\tilde{\w}_2) + (\tilde{\w}_3 \bar{d}\tilde{\w}_1 - \tilde{\w}_1 d\tilde{\w}_3) \\
&+ (\tilde{\w}_1 d\tilde{\w}_2 - \tilde{\w}_2 \bar{d}\tilde{\w}_1)	\\
=& -\pi^2+\pi^2-\pi^2	\\
=& -\pi^2.
\end{split}
\end{equation}
Substituting these results back into the line integral, we obtain
\begin{equation}
\begin{split}
\langle V_{111}(\gamma_1) V_{111}(\gamma_2) \rangle =& 4 \exp(\fr{i}{2\pi}\oint_{\gamma_1+\gamma_2}\sum_{ijk}\epsilon^{ijk} \bar{\w}_i d \bar{\w}_j)	\\
=& -4\punc{.}
\end{split}
\end{equation}
The correlation function suggests that the topological spin of $V_{111}$ is either $\ii$ or $-\ii$. A similar calculation for the linking correlation $\langle V_{n_1n_2n_3}(\gamma_1) V_{m_1m_2m_3}(\gamma_2) \rangle$ for type III twisted $\mbb{Z}_N^{\otimes3}$ theory can be found in Sec.~\ref{sec. ZN examples}. And we provide more comments on the topological spins there, and find the topological spin can actually fixed to be $-\ii$ for $V_{111}$, and hence $\ii$ for $V_{111}U_{111}$

In summary for this section (and App.~\ref{app. path integral}) we have calculated the correlation functions of linked line operators. The vanishing correlation functions (App.~\ref{app. path integral}) are further indications that type III twisted $\mbb{Z}_2^{\otimes3}$ is a non-abelian topological theory. The modular matrices of type III twisted $\mbb{Z}_2^{\otimes3}$ are explicitly written down in App.~\ref{app. ModMat}.

\section{Type III Twisted $\mbb{Z}_N^{\otimes3}$ Theory}
\label{sec. ZN examples}

In this section, we generalize the gauge group from $\mathbb{Z}_2^{\otimes 3}$ to $\mathbb{Z}_N^{\otimes 3}$. More explicitly, we construct the line operators and their correlation functions etc for the type III twisted $\mbb{Z}_N^{\otimes3}$ theories in (2+1)D. The basic idea of the constructing these line operators is still introducing the auxiliary fields and gauge invariance. Once we obtain the valid line operators, we can obtain their linking correlation function by path integral.

This section is divided into the following: In Sec.~\ref{app.sec. lag}, we again introduce the Lagrangian and gauge transformations; In Sec.~\ref{app.sec oper}, we list our line operators; In Sec.~\ref{app.sec. correlation}, we work out correlation functions of flux insertion operators. 

Moreover, two appendices are associated with this section: In App.~\ref{app.sec. gauge}, we verify the gauge invariance of flux insertion operators with lattice regularization; And in App.~\ref{app. QD}, we provide with a quantum double calculation which gives the same results of correlation functions as in the field theory approach derived in the following main text.

\subsection{Lagrangian and Gauge Transformation}
\label{app.sec. lag}

In this section, we introduce the Lagrangian for the twisted $\mbb{Z}_N^{\otimes3}$ theory and its gauge transformation, as a preparation for the following sections. The Lagrangian for the theory is
\begin{equation}\label{lag. 2+1 ZN}
\lag=\frac{N}{2\pi} b_i dA_i + \frac{pN^2}{(2\pi)^2} A_1 A_2 A_3,
\end{equation}
where $p\in\mbb{Z}_N=\{0,1,2,\ldots,N-1\}$, which can be determined by the same method in App.~\ref{app. coefficient}. The gauge transformations, by Eq.~\eqref{eq: gauge transf 2+1}, are
\begin{equation}\label{eq. gauge transf 2+1 ZN}
\begin{split}
b_i &\rightarrow b_i+d\beta_i+\fr{pN\epsilon^{ijk}}{2\pi}(A_j\alpha_k-\fr{1}{2}\alpha_jd\alpha_k)	\\
A_i &\rightarrow A_i+d\alpha_i,	\;\;	i=1,2,3.
\end{split}
\end{equation}
In the following sections, we will find out all line operators that are gauge invariant under the gauge transformation Eq.~\eqref{eq. gauge transf 2+1 ZN}, and work out some of their fusion rules and correlation functions etc.

\subsection{Line Operators}
\label{app.sec oper}

In this section, we find out all possible gauge invariant line operators. First of all, by gauge invariance, the Wilson line operators are:
\begin{equation}
U_{n_1n_2n_3}=\exp(\ii \oint_{\gamma} n_iA_i),	n_i\in\{0,1,.\ldots,N-1\}.
\end{equation}

Similar to the type III twisted $\mathbb{Z}_2^{\otimes 3}$ example in Sec.~\ref{sec: 2+1D}, the flux insertion operators can be constructed by introducing the auxiliary fields, $\phi_i$ and $\lambda_i$. From the decorated domain wall picture, the single type flux can be inferred from SPTs boundary with symmetry group $\mbb{Z}_{N}^{\otimes 2}$ instead of $\mbb{Z}_N^{\otimes 3}$. The reason is that in the path integral with a single flux insertion operator, for example $V_{100}$, we have flat connections $A_2$ and $A_3$ and non-flat connection $A_1$. Thus, the flux sheet of $A_1$ is actually a $\mathbb{Z}_N^{\otimes 2}$ SPT whose boundary is the flux loop. Therefore, by gauge anomaly inflow, we can construct the operator $V_{100}$ explicitly where auxiliary fields need to be introduced.

The auxiliary fields can be integrated out to produce a closed form for these flux insertion operators. However, instead of writing them down directly, we explain from the single type flux insertion operators to the triple type fluxes, mainly because the quantum dimensions of these operators are not written in a uniform way. 

The single type of flux insertion operators $V_{r00}$ are:
\begin{widetext}
\begin{equation}
\label{eq.r00}
\begin{split}
V_{r00}=&\fr{1}{\mathcal{N}}\int D[\phi_2]D[\phi_3]D[\lambda_2]D[\lambda_3] \exp(\ii \oint_{\gamma} rb_1+\fr{\ii rpN}{2\pi}\oint_{\gamma}\epsilon^{1ij}(\frac{1}{2}\phi_i d\phi_j + (d\phi_i - A_i)\lambda_j ))	\\
=& \mathcal{N}_{r00} \exp(\ii \oint_{\gamma} rb_1+\fr{\ii rpN}{4\pi}\oint_{\gamma}\epsilon^{1ij}\w_id\w_j) \delta(rp\bar{\w}_2)\delta(rp\bar{\w}_3),
\;\; r=0,1,\ldots, N-1.
\end{split}
\end{equation}
\end{widetext}
It is clearly gauge invariant as the $\phi_i$, $\lambda_i$ fields transform exactly the same as in $\mathbb{Z}_2^{\otimes 3}$ case. And the integration over the auxiliary fields are also the same. We emphasis that delta function $\delta(x)$ is still a projector, imposing any element $x \in 2\pi \mathbb{Z}$. The subtle difference from $\mathbb{Z}_2^{\otimes 3}$ is the normalization constant $\mathcal{N}_{r00}$ which is determined by fusion rules, for example $V_{r00} \times V_{(N-r)00}$. The flux is trivial after fusion. Hence we only expect charges appear in the fusion channels if the fusion is possibly nontrivial.

$\mathcal{N}_{r00}$ is fixed to be:
\begin{equation}\label{eq.r00 norm}
\mathcal{N}_{r00}=\frac{N}{\mathrm{gcd}(N,rp)}	\;\;.
\end{equation}
The reason for it is that the fusion rule of $V_{r00}$ and $V_{(N-r)00}$ is with Eq.~\eqref{eq.r00 norm}:
\begin{equation}\label{eq: fusion rule ZN}
V_{r00}\otimes V_{(N-r)00}=\bigoplus_{i,j=0}^{{\mathcal{N}_{r00}}-1} U_{0(ipr)(jpr)},
\end{equation}
where the fusion channels on the RHS has the greatest common divisor $1$, and the identity operator $U_{000}$ only appears once. The fusion rule is derived by taking the product of $V_{r00}$ and $V_{(N-r)00}$, canceling the exponential phases and expanding the $\delta$ function as follows:
\begin{equation}\label{eq. ZN delta}
\delta(rp\bar{\w}_j) = \frac{\mathrm{gcd}(N,rp)}{N} \sum\limits_{m=0}^{\frac{N}{\mathrm{gcd}(N,rp)}-1} \exp\left(\ii m rp \bar{\w}_j \right),
\end{equation}
where $j=2,3$. And we also use the fact for the derivation of fusion rule Eq.~\eqref{eq: fusion rule ZN}:
\begin{equation}
\mathcal{N}_{r00} = \frac{N}{\mathrm{gcd}(N,rp)} = \frac{N}{\mathrm{gcd}(N,(N-r)p)} = \mathcal{N}_{(N-r)00}.
\end{equation}

For the $\mathbb{Z}_2^{\otimes 3}$ example discussed in the previous section, where $N=2,p=1$, we have:
\begin{equation}
\mathcal{N}_{100}=\frac{2}{\mathrm{gcd}(2,1)}=2	\;.
\end{equation}

Having fixed the normalization for $V_{r00}$, the quantum dimensions for these flux insertion operators are just $\mathcal{N}_{r00}$, which can be manifested by calculating $\langle V_{r00} \rangle$. (More rigorously, the quantum dimension is obtained via fusion rules.) The explicit calculation is omitted here since it is exactly the same as in $\mathbb{Z}_2^{\otimes 3}$ situation.

Other types of single flux insertion operators $V_{0r0}$ and $V_{00r}$ can be obtained by simply permuting the indices, as in the previous section, Sec.~\ref{sec: 2+1D}. Hence we omit their expressions here for simplicity.

One can also consider inserting two types of fluxes and three types of fluxes. We follow the same prescription as in Eq.~\eqref{eq.111} by introducing the auxiliary fields $\phi_i$ and $\lambda_i$. And integrating the auxiliary fields out yields a closed form of flux insertion operators in terms of the holonomy functions $\w_i$'s:
\begin{widetext}
\begin{equation}
\begin{split}
V_{n_1n_20}=&\fr{1}{\mathcal{N}'}\int D[\phi_i]D[\lambda_i] \exp\left( \ii\oint_{\gamma} n_1 b_1+n_2 b_2+\fr{n_1pN}{2\pi}\epsilon^{1ij}(\frac{1}{2}\phi_i d\phi_j + (d\phi_i-A_i)\lambda_j) + \fr{n_2pN}{2\pi}\epsilon^{2ij}(\frac{1}{2}\phi_i d\phi_j + (d\phi_i-A_i)\lambda_j) \right)	\\
=& \mathcal{N}_{n_1n_20}\exp\left( \ii\oint_{\gamma} n_1 b_1 + n_2 b_2 - \fr{\ii Np}{2\pi}\oint_{\gamma}(n_2\w_1-n_1\w_2)d\w_3 \right) \delta(pn_2\bar{\w}_1-pn_1\bar{\w}_2)
\delta(n_1p\bar{\w}_3)\delta(n_2p\bar{\w}_3)
\end{split}
\end{equation}
\begin{equation}\label{eq. n1n2n3}
\begin{split}
V_{n_1n_2n_3}=&\fr{1}{\mathcal{N}''}\int D[\phi_i]D[\lambda_i] \exp\left(\ii\oint_{\gamma} (n_i b_i +\fr{n_ipN}{2\pi}\epsilon^{ijk}(\frac{1}{2}\phi_j d\phi_k + (d\phi_j-A_j)\lambda_k))\right)	\\
=& \mathcal{N}_{n_1n_2n_3} \exp\left(\ii\oint_{\gamma} (n_i b_i + n_i\frac{Np}{4\pi}\epsilon^{ijk}\w_jd\w_k)\right) \delta(n_2p\bar{\w}_1-n_1p\bar{\w}_2)\delta(n_3p\bar{\w}_2-n_2p\bar{\w}_3)\delta(n_1p\bar{\w}_3-n_3p\bar{\w}_1),
\end{split}
\end{equation}
\end{widetext}
where in Eq.~\eqref{eq. n1n2n3}, only two of the three $\delta$ functions are independent;  and $\mathcal{N}_{rs0}$ and $\mathcal{N}_{n_1n_2n_3}$ can be determined similarly as in $\mathcal{N}_{r00}$. More explicitly:
\begin{equation}\label{eq.rs0 norm}
\begin{split}
\mathcal{N}_{n_1n_20} &= \frac{N}{\mathrm{gcd}(N,pn_1,pn_2)}	\;\;,	\\
\mathcal{N}_{n_1n_2n_3}&=\frac{N}{\mathrm{gcd}(N,pn_1,pn_2,pn_3)}	\;\;.
\end{split}
\end{equation}
And similarly as before, the quantum dimensions of $V_{n_1n_20}$ and $V_{n_1n_2n_3}$ are $\mathcal{N}_{n_1n_20}$ and $\mathcal{N}_{n_1n_2n_3}$ respectively. They are also consistent with the case of $N=2, p=1$.  We write them in a uniform way:
\begin{equation}
\langle V_{n_1n_2n_3} \rangle = \mathcal{N}_{n_1n_2n_3}	\;.
\end{equation}

In the following, we will use the natural convention:
\begin{equation}
\begin{split}
\mathrm{gcd}(a,b,0,0) &\equiv \mathrm{gcd}(a,b)	\\
\mathrm{gcd}(a,b,c,0) &\equiv \mathrm{gcd}(a,b,c)
\end{split}
\end{equation}
to simplify our notations and discussions below. We can write all the flux insertion operators uniformally by using this notation. 

In summary, we have determined the flux insertion operators as in Eq.~\eqref{eq. n1n2n3}. They are constructed by introducing the auxiliary fields on the fluxes, and integrating out the auxiliary fields yields the closed forms for these operators in terms of the holonomy functions $\w_i$. The quantum dimensions for the flux insertion operators are determined by the overall coefficients in the closed form of flux insertion operators. See Eq.~\eqref{eq.r00 norm} and \eqref{eq.rs0 norm}. Moreover, in App.~\ref{app.sec. gauge}, we verify the gauge invariance of these operators Eq.~\eqref{eq. n1n2n3} explicitly with lattice regularizations.

We do not elaborate on how many different line operators here, but only comment that because of the $\delta$ functions in the flux insertion operators $V_{n_1n_2n_3}$, attaching a Wilson line onto $V_{n_1n_2n_3}$ may actually contribute nothing trivial phases to the correlation functions. Hence, some operators are identified in the sense of producing the same correlation functions, although their appearances are different.

\subsection{Correlation Functions}
\label{app.sec. correlation}

In this section, we provide with general linking correlation functions for two flux insertion operators, $\langle V_{n_1n_2n_3}(\gamma_1)V_{m_1m_2m_3}(\gamma_2) \rangle$, for the type III twisted $\mbb{Z}_N^{\otimes3}$ theory.  Before we calculate $\langle V_{n_1n_2n_3}(\gamma_1)V_{m_1m_2m_3}(\gamma_2) \rangle$, we first comment on other simpler linking correlation function, for example the linking of two Wilson lines, or Wilson lines and flux insertion operators.

The linking correlations between any two Wilson lines are simply identity. And the linking correlations between flux insertion operators and Wilson line operators remain to be simple. Flux insertion operator $V_{n_1n_2n_3}$ simply inserts $n_1$ units of $A_1$ flux, $n_2$ units of $A_2$ flux and $n_3$ units of $A_3$ flux. Thus Wilson lines that are linked to $V_{n_1n_2n_3}$ simply take three Aharonov-Bhom phases according to the charges of the Wilson lines. However, as we have seen in $\mathbb{Z}_2^{\otimes 3}$ section, the linking correlation functions may vanish due to the constraint part of these operators, or pick up nontrivial phases from the $\w_i d\w_j$ terms.

In the following formulas, we do not distinguish the lattice derivatives $d$ and $\bar{d}$ until necessary. And more importantly, we assume $\mathrm{gcd}(N,p)=1$ which simplifies the calculations. The explanations for the assumption will be explained after the calculations. The detailed calculation goes as follows:
\begin{widetext}
\begin{equation}\label{eq. ZN linking}
\begin{split}
&\langle V_{n_1n_2n_3}(\gamma_1) V_{m_1m_2m_3}(\gamma_2) \rangle	\\
=& \mathcal{N}_{n_1n_2n_3}\mathcal{N}_{m_1m_2m_3} \int Db_i DA_i \exp\left(\ii \int \frac{N}{2\pi} b_i dA_i + \frac{pN^2}{(2\pi)^2} A_1 A_2 A_3 \right)	\\
&\exp\left(\ii\oint_{\gamma_1}n_ib_i+\fr{\ii Np}{4\pi}\oint_{\gamma_1}\epsilon^{ijk} n_i \w_j d \w_k\right) \prod_i \delta(\epsilon^{ijk}pn_i\bar{\w}_k|_{\gamma_1})	\\
&\exp\left(\ii\oint_{\gamma_2}m_ib_i+\fr{\ii Np}{4\pi}\oint_{\gamma_2}\epsilon^{ijk} m_i \w_j d \w_k\right) \prod_i \delta(\epsilon^{ijk}pm_i\bar{\w}_k|_{\gamma_2})	\\
=& \mathcal{N}_{n_1n_2n_3}\mathcal{N}_{m_1m_2m_3} \exp\left(\ii\frac{pN^2}{(2\pi)^2}\int\tilde{A}_1\tilde{A}_2\tilde{A}_3\right)
\exp\left(\fr{\ii Np}{4\pi}\oint_{\gamma_1}\epsilon^{ijk} n_i \tilde{\w}_j d \tilde{\w}_k\right)
\exp\left(\fr{\ii Np}{4\pi}\oint_{\gamma_2}\epsilon^{ijk} m_i \tilde{\w}_j d \tilde{\w}_k\right)	\\
&\prod_i \delta(\epsilon^{ijk}pn_j\frac{2\pi m_k}{N}) \prod_i \delta(\epsilon^{ijk}pm_j\frac{2\pi n_k}{N})	\\
=& \mathcal{N}_{n_1n_2n_3}\mathcal{N}_{m_1m_2m_3} \exp\left( \fr{\ii Np}{4\pi}\oint_{\gamma_1}\epsilon^{ijk} n_i \tilde{\w}_j d \tilde{\w}_k\right)	\exp\left( \fr{\ii Np}{4\pi}\oint_{\gamma_2}\epsilon^{ijk} m_i \tilde{\w}_j d \tilde{\w}_k\right)	\prod_i \delta(\epsilon^{ijk}\frac{2\pi p}{N} n_jm_k)\prod_i \delta(\epsilon^{ijk}\frac{2\pi p}{N} m_jn_k)	\\
=& \mathcal{N}_{n_1n_2n_3}\mathcal{N}_{m_1m_2m_3}
\exp\left(-\fr{\ii p\pi}{N}\epsilon^{ijk}(n_im_jm_k + m_in_jn_k)\right)	
\prod_i \delta(\epsilon^{ijk}\frac{2\pi p}{N} n_jm_k) \prod_i \delta(\epsilon^{ijk}\frac{2\pi p}{N} m_jn_k).
\end{split}
\end{equation}
\end{widetext}
The equations deserves certain explanations: The first equality just lists all the terms, following the same convention as before. In the second equality, we integrate out all $b_i$ fields, which yields the equations of motion as follows:
\begin{equation}\label{eq. ZN eom}
dA_i = \frac{2\pi}{N}(n_i\star j_1 + m_i\star j_2) \mod{2\pi},~i=1,2,3,
\end{equation}
where $j_1$ and $j_2$ are the currents representing $\gamma_1$ and $\gamma_2$ respectively, $\star$ is the Hodge dual. The solutions of such equations of motion are denoted as $\tilde{A}_i,~i=1,2,3$.
As a result of the equations of motion, we have:
\begin{equation}
\begin{split}
\bar{\w}_i(\gamma_1)&\equiv\oint_{\gamma_1}A_i=\frac{2\pi}{N} m_i,	\\
\bar{\w}_i(\gamma_2)&\equiv\oint_{\gamma_2}A_i=\frac{2\pi}{N} n_i,~i=1,2,3.
\end{split}
\end{equation}
Now notice that due to the $\delta$ function constraints, the correlation will vanish if they are not satisfied. One subtlety is that the $\delta$ functions associated with $V_{n_1n_2n_3}$ and $V_{m_1m_2m_3}$ are slightly different, because the summation periods as in Eq.~\eqref{eq. ZN delta} are determined by $\mathcal{N}_{n_1n_2n_3}$ and $\mathcal{N}_{m_1m_2m_3}$ respectively:
\begin{equation}
\begin{split}
& \delta(\epsilon^{ijk}\frac{2\pi p}{N}n_jm_k) 	\\
=& \frac{1}{\mathcal{N}_{n_1n_2n_3}} \sum\limits_{q=0}^{\mathcal{N}_{n_1n_2n_3}-1} \exp\left(\ii q \epsilon^{ijk}\frac{2\pi p}{N}n_jm_k\right)	\\
& \delta(\epsilon^{ijk}\frac{2\pi p}{N}m_jn_k)	\\
=& \frac{1}{\mathcal{N}_{m_1m_2m_3}} \sum\limits_{q=0}^{\mathcal{N}_{m_1m_2m_3}-1} \exp\left(\ii q \epsilon^{ijk}\frac{2\pi p}{N}m_jn_k\right).
\end{split}
\end{equation}
These $\delta$ functions leads to the following equations:
\begin{equation}\label{eq. constraint result}
\begin{split}
& n_2m_3-n_3m_3	= 0 \mod{\frac{N}{\mathrm{gcd}(N,p)}},	\\
& \text{and permuted equations.}
\end{split}
\end{equation}

For the convenience of the following calculations, we assume that $\mathrm{gcd}(N,p)=1$. Thus the above equations Eq.~\eqref{eq. constraint result} are valid mod $N$. Therefore, the RHS of the equations of motions Eq.~\eqref{eq. ZN eom} are proportional to each other, for $i=1,2,3$. Hence, the solutions, $\tilde{A}_i,~i=1,2,3$, can be set to proportional to each other, up to gauge transformations. Hence, the first integral of the second equality will vanish at least in the continuous limit, leading to the third equality. Note that if $\mathrm{gcd}(N,p)\neq 1$, the argument that $\int \tilde{A}_1\tilde{A}_2\tilde{A}_3$ vanishes may not be true. 

The fourth equality is obtained by using the lattice derivatives, $d$ and $\bar{d}$. For example:
\begin{equation}
\oint_{\gamma_1} \tilde{\w}_2 d\tilde{\w}_3 - \tilde{\w}_3 \bar{d} \tilde{\w}_2 = -\left(\frac{2\pi}{N}\right)^2 m_2m_3.
\end{equation}
Thus, we have completed the calculation of $\langle V_{n_1n_2n_3}(\gamma_1)V_{m_1m_2m_3}(\gamma_2) \rangle$.

One particular simple and non-vanishing example of these linking correlations is
\begin{equation}\label{eq. linking n1n2n3}
\begin{split}
&\langle V_{n_1n_2n_3}(\gamma_1) V_{n_1n_2n_3}(\gamma_2)\rangle	\\
=&\mathcal{N}_{n_1n_2n_3}^2 \exp\left(\frac{-2\pi\ii p}{N}n_1n_2n_3\right)	\;.
\end{split}
\end{equation}
where we need to recall that:
\begin{equation}
\mathcal{N}_{n_1n_2n_3} = \frac{N}{\mathrm{gcd}(N,pn_1,pn_2,pn_3)}.
\end{equation}

\begin{figure}[t]
\centering
\includegraphics[width=0.3\textwidth]{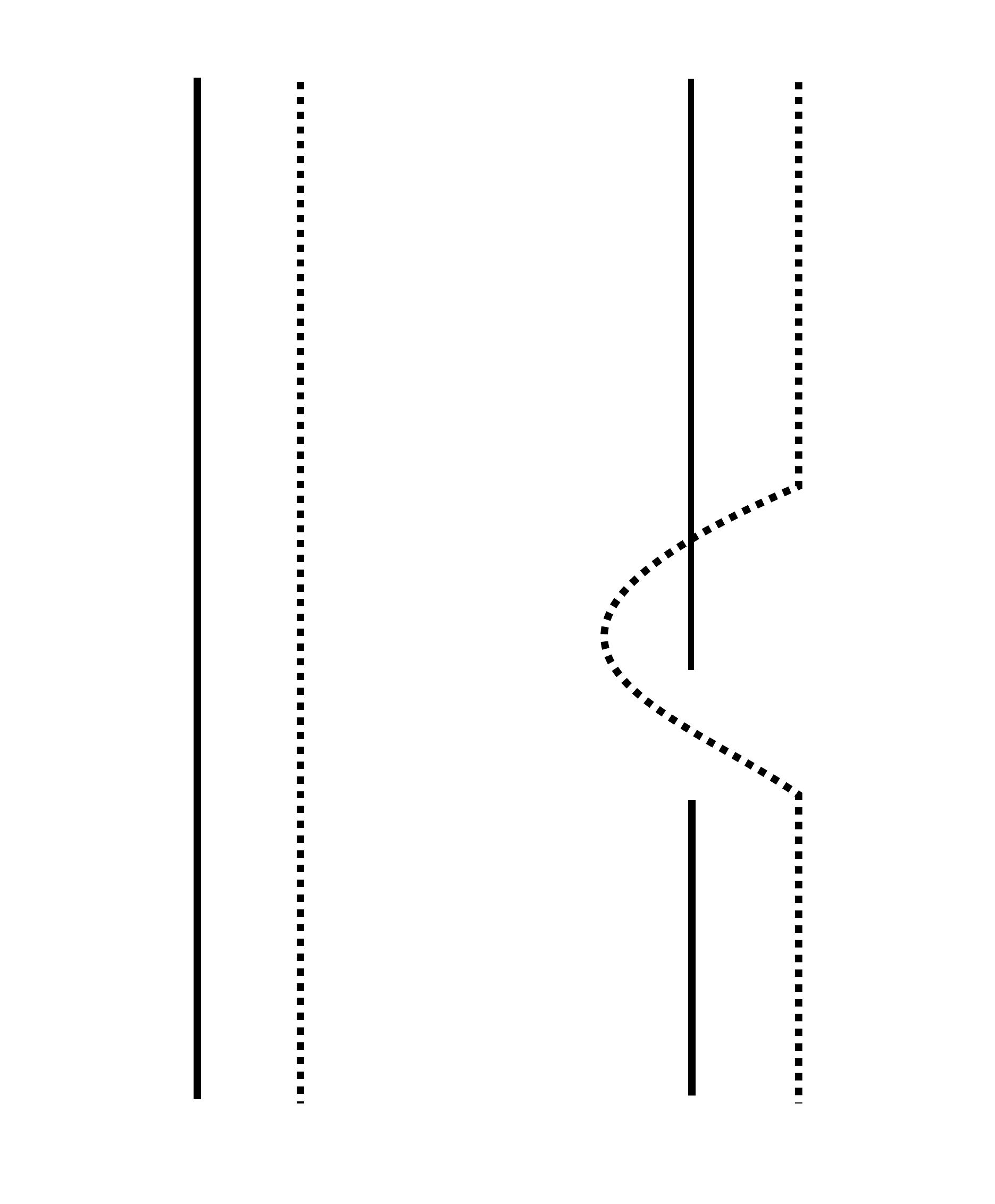}
\caption{An illustration of self-twist. The left panel is a ribbon without self-twists. The dotted line is the illustration of framing for the solid line. The right panel is a ribbon with one self-twist. The dotted line, the framing, winds around the solid line once\cite{witten1989}. We can simply view the right panel as a link of the solid and dotted line.}
\label{fig.ribbon}
\end{figure}

The linking correlation function Eq.~\eqref{eq. linking n1n2n3} also suggests the topological spin for $V_{n_1n_2n_3}$ is:
\begin{equation}\label{eq. spin ZN}
\Theta(V_{n_1n_2n_3}) = \exp\left( -\frac{\pi\ii p}{N}n_1n_2n_3 \right)	\;,
\end{equation}
although it is a non-abelian topological phase. The reason is the following: Suppose we have a ``self-twist" loop $\gamma$. See Fig.~\ref{fig.ribbon} for an illustration of self-twist. Then $V_{n_1n_2n_3}(\gamma)$ itself inserts $n_{1,2,3}$ units of $A_{1,2,3}$ fluxes through $\gamma$. Hence in the path integral of $\langle V_{n_1n_2n_3}(\gamma) \rangle$, we only have one contribution for the phase, instead of two contributions as in the last two equality of Eq.~\eqref{eq. ZN linking}. The $\delta$ functions in the $V_{n_1n_2n_3}$ are automatically satisfied and hence do not contribute.

Restricting to the $\mathbb{Z}_2^{\otimes 3}$ situation where $N=1,p=1$, we actually have $\Theta(V_{111}) = -\ii$\cite{propitius1995topological}. In terms of quantum double notations in Sec.~\ref{subsec: fusion rules 2+1}, $\Theta((111,\gamma_{+}))=-\ii$. Moreover, when the Wilson line contributing a minus sign in the path integral is attached to $V_{111}$, the topological spin obtains one more minus sign. Hence, we find that $\Theta((111,\gamma_{-}))=\ii$\cite{propitius1995topological}.

In App.~\ref{app. QD}, we provide a quantum double calculation and calculate the projective representations determined by the slant product of type III cocycles with group $\mathbb{Z}_N^{\otimes 3}$. And we show that it gives the same correlation functions etc as by the field theoretical approach above.

In summary for this section, we calculated the linking correlation functions of flux insertion operators explicitly, Eq~\eqref{eq. ZN linking}, with an assumption $\mathrm{gcd}(N,p)=1$. And as a consequence, we obtain the topological spins for the flux insertion operators, Eq~\eqref{eq. spin ZN}.

\section{Conclusion}
In this work we considered a continuum formulation of abelian Dijkgraaf-Witten field theories in (2+1)D. These theories come in three varieties: types I, II and III. We constructed all the possible gauge invariant line operators, which correspond to the possible quasi-particle excitations. The quasi-particles of type I and type II theories are readily understood using a $K$-matrix Chern-Simons theory approach. We mostly consider the subtler issue of type III DW models focussing on type III twisted $\mbb{Z}_2^{\otimes 3}$ and more generally $\mbb{Z}_N^{\otimes 3}$ DW theory. Despite the fact these theories have abelian gauge groups, their excitations have non-abelian fusions and statistics. We demonstrated this by directly constructing all Wilson and flux insertion operators, and computing  all of their associated braiding and fusion rules. The guiding principle in constructing these operators is gauge invariance which, once imposed, leads to the introduction of auxiliary fields which live on the line operators in question. These auxiliary fields can be viewed as internal degrees of freedom of the particle in question. 

Our work thus provides a field-theoretical platform for analyzing non-abelian (2+1)D SPTs and topological orders. It would be useful to extend some of the constructions here to higher dimensions, where topological phases are less well understood\cite{moradi2015,wang2015non,juvenwang2016quantum,wang2014,wang2015,lin2015,wang2016,jiang2014,wan2015twisted}.

\acknowledgements
HH and YZ thank B. A. Bernevig, E. Witten for useful comments and discussions.

CvK acknowledges the support of the Princeton Center for Theoretical Science.
HH and YZ acknowledge the support from Department of Physics, Princeton University.

\appendix

\newpage

\section{Classification of Twist Terms}
\label{app. coefficient}

In this appendix, we will basically repeat the main idea of Ref.~\onlinecite{wang2015field}, for the purpose of completeness for this work. We will explain how to fix the coefficients of the twist terms in the Lagrangian in (2+1)D. Of course, the method can be generalized to other types of twists, other gauge groups, and other dimensions. For details, please refer to Ref.~\onlinecite{wang2015field}.

We check our results by noting that in (2+1)D, the spectra of DW models are described algebraically as quasi-quantum double models twisted by cocycles\cite{propitius1995topological} (see also Ref.~\onlinecite{wan2015twisted,hu2013twisted}).

Presumably, the fields in the Lagrangian Eq.~\eqref{eq: lag 2+1} are $\mbb{Z}_2$ variables valued in $\{0,\pi\}$. And their holonomies satisfy:
\begin{equation}
\oint A_i = n_i\pi,~n_i\in\mbb{Z},~\forall i.
\end{equation}

In order to fix the coefficient of $A_1A_2A_3$. We need two requirements: invariance under large gauge transformation, and flux identification. The generator of the large gauge transformation is defined as:
\begin{equation}
\oint A_i \mapsto \oint A_i + 2\pi,~\forall i
\end{equation}

\paragraph{Large Gauge Transformation}: Under large gauge transformation, supposing only to $A_1$, then $\int A_1A_2A_3 \mapsto \int A_1A_2A_3 + \int \delta A_1 A_2 A_3 = \int A_1A_2A_3 + 2\pi^3 n_2n_3$. The invariance of the action Eq.~\eqref{eq: lag 2+1} under large gauge transformation gives:
\begin{equation}
\frac{pn_2n_3}{4}\in\mbb{Z}.
\end{equation}
Symmetrically, we have
\begin{equation}
\frac{pn_1n_2}{4},\frac{pn_1n_3}{4}\in\mbb{Z}.
\end{equation}

For arbitrary integers $n_1$, $n_2$ and $n_3$, we have
\begin{equation}
p\in 4\mbb{Z}.
\end{equation}

\paragraph{Flux Identification}: The integral of $A_1A_2A_3$ term is actually
\begin{equation}
\frac{p}{(2\pi)^2} \int A_1 A_2 A_3 = \frac{p}{(2\pi)^2} n_1n_2n_3 \pi^3=\frac{pn_1n_2n_3}{4}\pi.
\end{equation}

So when $p$ is shifted to $p+8$, the integral does not change, which implies that $p$ should be identified with $p+8$.

\paragraph{Summary}: Combining the two requirements, we conclude that $p$ is valued in $\{0,4\}$. In the main text, we simply choose the nontrivial value of $p$:
\begin{equation}
\lag=\frac{2}{2\pi} b_i dA_i + \frac{4}{(2\pi)^2} A_1 A_2 A_3	\;.
\end{equation}

Generalization from group $\mathbb{Z}_2$ to group $\mathbb{Z}_N$ is direct. The holonomies are quantized to:
\begin{equation}
\oint A_i = \frac{2\pi}{N} n_i,~n_i\in\mbb{Z},~\forall i.
\end{equation}
And large gauge transformations remain to be:
\begin{eqnarray}
\oint A_i \mapsto \oint A_i + 2\pi,~\forall i
\end{eqnarray}
Repeat the same calculation, we can fixed the coefficients of type III twisted $\mathbb{Z}_N^{\otimes 3}$ to be:
\begin{equation}
\lag=\frac{N}{2\pi} b_i dA_i + \frac{pN^2}{(2\pi)^2} A_1 A_2 A_3	\;,
\end{equation}
where $p\in\mbb{Z}_N=\{0,1,2,\ldots,N-1\}$.

\section{Details of Calculating the Path Integral in (2+1)D}
\label{app. path integral}

The path integrals of DW models can be rigorously calculated especially when they are regulated on lattice. In this section, we will explain the methodology by doing two examples of path integral calculation which we have constantly been using in this work. 

The rest of the appendix is divided into two parts: In the first one, we derive the closed form of flux insertion operators by integrating out the auxiliary fields; in the second part, we show two correlation functions as we promised in our main text. The first correlation is simple $\langle V_{100} \rangle$, and the second one is a linking correlation $\langle V_{100}(\gamma_1) V_{010}(\gamma_2) \rangle$. Both correlation functions suggest that the theory is actually the non-Abelian topologically ordered: $\langle V_{100} \rangle$ is the quantum dimension for the operator $V_{100}$ which is larger than 1, while $\langle V_{100}(\gamma_1) V_{010}(\gamma_2) \rangle$ vanishes other than a $U(1)$ phase.

\subsection{Closed Form of flux insertion Operators}

We begin with $V_{100}$ operator in Eq.~\eqref{eq.100}, and show how the closed form of the $V_{100}$ is deduced. Suppose the calculation is well-regulated on lattice, and $\gamma$ is a closed line with $L$ bonds. The variables in the functional integral are $\mbb{Z}_2$ variables valued in $\{0,\pi\}$.

The calculation details goes as follows:
\begin{widetext}
\begin{equation}
\begin{split}
V_{100}(\gamma)=&\frac{1}{\mathcal{N}} \int D[\phi_2]D[\phi_3]D[\lambda_2]D[\lambda_3]	\exp(\ii\oint_{\gamma} b_1 + \frac{\epsilon^{1ij}}{\pi} (\frac{1}{2} \phi_i d \phi_j + (d\phi_i - A_i) \lambda_j ))	\\
=&\frac{2^{2L}}{\mathcal{N}} \int D[\phi_2]D[\phi_3] \exp(\ii\oint_{\gamma} b_1 + \frac{1}{\pi}\phi_2 d \phi_3) \delta(d\phi_2-A_2)\delta(d\phi_3-A_3)	\\
=&\frac{2^{2L}}{\mathcal{N}} \int D[\phi_2]D[\phi_3] \exp(\ii\oint_{\gamma} b_1 + \frac{1}{\pi}\phi_2 d \phi_3) \delta(\phi_2-\w_2 - C_2)\delta(\phi_3-\w_3 - C_3)	\\
=&\frac{2^{2L}}{\mathcal{N}} \exp(\ii\oint_{\gamma} b_1 + \frac{1}{\pi}\w_2 d \w_3) \delta(\bar{\w}_2)\delta(\bar{\w}_3)	\\
=& 2\exp(\ii\oint_{\gamma} b_1 + \frac{1}{\pi}\w_2 d \w_3) \delta(\bar{\w}_2)\delta(\bar{\w}_3).
\end{split}
\end{equation}
\end{widetext}
The notations in the above equations include: $\w_2$ and $\w_3$ are holonomy function $w_2=\int_{0}^{x} A_2$, $w_3=\int_{0}^{x} A_3$. And $\bar{\w}_2=\int_{0}^{L} A_2$ and $\bar{\w}_3=\int_{0}^{L} A_3$. Note that we need the constraint $\delta(\bar{\w_2})\delta(\bar{\w}_3)$ in order to define $V_{100}$, otherwise it is not gauge invariant. The above calculation deserves certain explanations: in the second equality, the $\lambda_2$ and $\lambda_3$ are actually Lagrangian multipliers and integrating them out yields two constraints; in the third equality, we just solve the constraints; the rest of the calculations are natural, except that the reason of choosing normalization factor $\mathcal{N}=2^{2L-1}$ is to have the coefficients of fusion rules integers. That was explained in the main text.

$V_{110}$ and $V_{111}$ can be deduced similarly. For completeness, we provide one more example, $V_{111}$, while $V_{110}$ is less subtle.

\begin{widetext}
\begin{equation}
\begin{split}
V_{111}=&\fr{1}{\mathcal{N}}\int D[\phi_i]D[\lambda_i] \exp(\ii\oint_{\gamma} b_1 + b_2 + b_3 + \frac{\epsilon^{ijk}}{\pi} (\frac{1}{2}\phi_j d \phi_k + (d\phi_j-A_j)\lambda_k))	\\
=&\fr{2^{3L}}{\mathcal{N}}\int D[\phi_i] \exp(\ii\oint_{\gamma} b_1 + b_2 + b_3 + \frac{\epsilon^{ijk}}{2\pi}\phi_j d \phi_k)	\\ &\delta(d\phi_1-d\phi_2-A_1+A_2)\delta(d\phi_2-d\phi_3-A_2+A_3)\delta(d\phi_3-d\phi_1-A_3+A_1)	\\
=&\fr{2^{3L}}{\mathcal{N}}\int D[\phi_i] \exp(\ii\oint_{\gamma} b_1 + b_2 + b_3 + \frac{\epsilon^{ijk}}{2\pi}\phi_j d \phi_k)	\\
& \delta(\phi_1-\w_1 - f_0 - C_1)\delta(\phi_2-\w_2 - f_0 - C_2)\delta(\phi_3-\w_3 - f_0 - C_3)	\\
=&\fr{2^{3L}}{\mathcal{N}} \exp(\ii\oint_{\gamma} b_1 + b_2 + b_3 + \frac{\epsilon^{ijk}}{2\pi}(\w_j + f_0) d (\w_k + f_0)) \delta(\bar{\w}_1-\bar{\w}_2)\delta(\bar{\w}_2-\bar{\w}_3)	\\
=&\fr{2^{3L}}{\mathcal{N}} \exp(\ii\oint_{\gamma} b_1 + b_2 + b_3 + \frac{\epsilon^{ijk}}{2\pi}\w_j d \w_k) \delta(\bar{\w}_1-\bar{\w}_2)\delta(\bar{\w}_2-\bar{\w}_3).
\end{split}
\end{equation}
\end{widetext}
The calculation is quite similar to $V_{100}$. The only thing that changes is the constraints by the $\delta$ functions. The solution of the constraints in the second equality is $\phi_2=\w_2-\w_1+\phi_1+C_2$ and $\phi_3=\w_3-\w_1+\phi_1+C_3$. A more symmetric way of expressing the same solutions are $\phi_i = \w_i - \w_0 + \phi_0 + C_i,~i=1,2,3$ by using a common ``reference" $\w_0$ and $\phi_0$. The constants $C_i$ can be shifted away. One subtlety needs our attention: the existence of the solutions requires that $\bar{\w}_1=\bar{\w}_2=\bar{\w}_3$. However, we choose a more symmetric way to express the solutions as in the third equality, which will simplify the expansion from the fourth equality to the fifth.

\subsection{Correlation Function}

As we promised in the main text, in this section, we will illustrate how to work out the correlation functions by doing two examples: the first one is $\langle V_{100} \rangle$; the second one is the correlation of $V_{100}(\gamma_1)$ and $V_{010}(\gamma_2)$ where $\gamma_1$ and $\gamma_2$ form a link in (2+1)D.

The correlation of a single $V_{100}(\gamma)$ is:
\begin{widetext}
\begin{equation}
\begin{split}
& \langle V_{100}(\gamma) \rangle	\\
=& \int D[b_i] D[A_i] \exp(\ii\int \frac{1}{\pi} b_i dA_i + \frac{1}{\pi^2} A_1A_2A_3)
2\exp(\ii\oint_{\gamma} b_1 + \frac{\epsilon^{1ij}}{2\pi} \w_i d \w_j)
\delta(\bar{\w}_2|_{\gamma})\delta(\bar{\w}_3|_{\gamma})	\\
=& 2\exp(\ii\int \frac{1}{\pi^2} \tilde{A}_1\tilde{A}_2\tilde{A}_3) \exp(\ii\oint_{\gamma}\frac{1}{\pi} \tilde{\w}_2 d \tilde{\w}_3) \delta(\bar{\w}_2|_{\gamma})\delta(\bar{\w}_3|_{\gamma})	\\
=& 2.
\end{split}
\end{equation}
\end{widetext}
We denote the $A_i,~i=1,2,3$ after integrating out $b_i,~i=1,2,3$ as $\tilde{A}_i,~i=1,2,3$. Integrating out $b_2$ and $b_3$ will yield a flat $A_2$ and $A_3$. And we choose the gauge $\tilde{A}_2=\tilde{A_3}=0$. So the phases in the right above equations will be trivial. And the $\delta$ functions are all satisfies: $\delta(\bar{\w}_2)=\delta(\bar{\w}_3)=1$. Integrating out $b_1$ will make $\tilde{A}_1$ have a $\pi$ flux surrounding $\gamma$. For other loops that do not surround $\gamma$, $\tilde{A}_1$ has a trivial flux.

The correlation of $V_{100}(\gamma_1)$ and $V_{010}(\gamma_2)$ goes as follows:
\begin{widetext}
\begin{equation}
\begin{split}
&\langle V_{100}(\gamma_1) V_{010}(\gamma_2) \rangle	\\
=& \int D[b_i] D[A_i] \exp(\ii\int \frac{1}{\pi} b_i dA_i + \frac{1}{\pi^2} A_1A_2A_3)
~2\exp(\ii\oint_{\gamma_1} b_1 + \frac{\epsilon^{1ij}}{2\pi} \w_i d \w_j) \delta(\bar{\w}_2|_{\gamma_1})\delta(\bar{\w}_3|_{\gamma_1})	\\
& 2\exp(\ii\oint_{\gamma_2} b_2 + \frac{\epsilon^{2ij}}{2\pi} \w_i d \w_j) \delta(\bar{\w}_1|_{\gamma_2})\delta(\bar{\w}_3|_{\gamma_2})	\\
=& 4\exp(\ii\int \frac{1}{\pi^2} \tilde{A}_1\tilde{A}_2\tilde{A}_3)	\exp(\ii\oint_{\gamma_1} \frac{1}{\pi} \tilde{\w}_2 d \tilde{\w}_3)
\exp(\ii\oint_{\gamma_2} \frac{1}{\pi} \tilde{\w}_3 d \tilde{\w}_1) \delta(\tilde{\bar{\w}}_1|_{\gamma_2})\delta(\tilde{\bar{\w}}_2|_{\gamma_1})\delta(\tilde{\bar{\w}}_3|_{\gamma_1})\delta(\tilde{\bar{\w}}_3|_{\gamma_2})	 \\
=& 4 \exp(\ii\oint_{\gamma_1} \frac{1}{\pi} \tilde{\w}_2 d \tilde{\w}_3) \exp(\ii\oint_{\gamma_2} \frac{1}{\pi} \tilde{\w}_3 d \tilde{\w}_1)	 \delta(\tilde{\bar{\w}}_1|_{\gamma_2})\delta(\tilde{\bar{\w}}_2|_{\gamma_1})\delta(\tilde{\bar{\w}}_3|_{\gamma_1})\delta(\tilde{\bar{\w}}_3|_{\gamma_2})	 \\
=& 4 \exp(\ii\oint_{\gamma_1} \frac{1}{\pi} \tilde{\w}_2 \tilde{A}_3) \exp(-\ii\oint_{\gamma_2} \frac{1}{\pi} \tilde{A}_3 \tilde{\w}_1)	 \delta(\tilde{\bar{\w}}_1|_{\gamma_2})\delta(\tilde{\bar{\w}}_2|_{\gamma_1})\delta(\tilde{\bar{\w}}_3|_{\gamma_1})\delta(\tilde{\bar{\w}}_3|_{\gamma_2})\\
=& 4\delta(\tilde{\bar{\w}}_1|_{\gamma_2})\delta(\tilde{\bar{\w}}_2|_{\gamma_1})\delta(\tilde{\bar{\w}}_3|_{\gamma_1})\delta(\tilde{\bar{\w}}_3|_{\gamma_2})	 \\
=& 0.
\end{split}
\end{equation}
\end{widetext}

Once we integrate out $b_1$ and $b_2$, $A_1$ will have a unit $\pi$ flux surrounding $\gamma_1$ and $A_2$ have a $\pi$ flux surrounding $\gamma_2$. Integrating out $b_3$ will produce a flat $A_3$. Suppose we choose the gauge orbit for $A_3=0$. The notation $\tilde{A}_i,~\tilde{\w}_i,~i=1,2,3$ are denoted for the fields after integrating out $b_i,~i=1,2,3$, and $\tilde{A}_3=0$. Then most of the phases in the calculation will end up being trivial. Therefore we only have four $\delta$ functions in the last two equality in the right above. Note that $\gamma_1$ and $\gamma_2$ form a link. So that $\gamma_1$ and $\gamma_2$ will surround each other. Integrating out $b_i$ will yield $\oint_{\gamma_2}\tilde{A_1}=\oint_{\gamma_1}\tilde{A_2}=\pi$. The $\delta$ function will be violated. Therefore, the correlation is simply $0$.

\section{Gauge Invariance of $V_{n_1n_2n_3}$ on Lattice}
\label{app.sec. gauge}

We could easily verify that the flux insertion operators $V_{n_1n_2n_3}$ is gauge invariant in the continuous limit. However, as we have utilized a lattice regularization in the main text to calculate the partition function, it is necessary to verify the operators $V_{n_1n_2n_3}$ are still gauge invariant on lattice. In this section, we will verify the gauge invariance explicitly for flux insertion operators on lattice.

Before we start, we need to do some basic mathematical preparations for the lattice derivatives and lattice integral. Note that in the gauge transformation of $b_1$, Eq.~\eqref{eq. gauge transf 2+1 ZN}, $A_{2,3}$ and $\alpha_{2,3}$ are coupled. We need to firstly specify how it is coupled on lattice. For convenience, we use our notations $\w_{2,3}$ and lattice derivatives $d,~\bar{d}$ to make the coupling obvious. For clearness, we repeat the definitions here although they have been mentioned in the main text. At $r-$th site, the lattice derivatives are defined as below:
\begin{equation}
\begin{split}
d\w_i(r) &= \w_i(r+1)-\w_i(r)	\\
\bar{d}\w_i(r) &= \w_i(r)-\w_i(r-1),\;\;i=1,2,3.
\end{split}
\end{equation}
Note that either $d$ or $\bar{d}$ is chosen, we still have
\begin{eqnarray}
\oint d\w_i = \oint \bar{d}\w_i = \bar{\w}_i,~i=1,2,3.
\end{eqnarray}

And we have the lattice version of integral by part for arbitrary functions $f$ and $g$ (they may not be periodic on the lattice of size $L$). We start with considering the following integral on lattice:
\begin{equation}
\begin{split}
&	\oint f d g + g \bar{d} f \\
=&	\sum_{r=0}^{L-1} f(r)(g(r+1)-g(r)) + \sum_{r=1}^{L} g(r)(f(r)-f(r-1))	\\
=&	f(L)g(L) - f(0)g(0).
\end{split}
\end{equation}

It can be put in a integral-by-part-theorem way:
\begin{equation}
\oint f d g = f(L)g(L) - f(0)g(0) - \oint g \bar{d} f.
\end{equation}

For $f$ and $g$ that are single valued on the lattice ring, the integral-by-part theorem reduces to:
\begin{equation}
\oint f d g = - \oint g \bar{d} f	\;\;.
\end{equation}

With these preparations of lattice derivatives and lattice integrals, we can start to verify the gauge invariance. First of all, we consider the gauge transformations for the $b_1$ related terms, $\oint b_1 + \frac{pN}{4\pi}(\w_2 d\w_3 - \w_3 \bar{d}\w_2)$, as follows:
\begin{widetext}
	\begin{equation}
	\begin{split}
	&\oint b_1 + \frac{pN}{4\pi}(\w_2 d\w_3 - \w_3 \bar{d}\w_2)	\Rightarrow	\\
	&\oint b_1 + \frac{pN}{4\pi}(2 \bar{d}\w_2 \alpha_3 - 2d\w_3 \alpha_2 - \alpha_2 d\alpha_3 + \alpha_3 \bar{d}\alpha_2 ) + \frac{pN}{4\pi}((\w_2 + \alpha_2 - \alpha_2(0)) (d\w_3 + d\alpha_3) - (\w_3 + \alpha_3 - \alpha_3(0)) (\bar{d}\w_2 + \bar{d}\alpha_2))	\\
	=&\oint b_1 + \frac{pN}{4\pi}(\w_2 d\w_3 - \w_3 \bar{d}\w_2) + \frac{pN}{2\pi}( \bar{d}\w_2 \alpha_3 - d\w_3 \alpha_2) + \frac{pN}{4\pi}[\w_2 d\alpha_3 + (\alpha_2 - \alpha_2(0))d\w_3 - \w_3 \bar{d}\alpha_2 - (\alpha_3 - \alpha_3(0))\bar{d}\w_2]	\\
	=&\frac{pN}{4\pi}(\alpha_3(0)\bar{\w}_2-\alpha_2(0)\bar{\w}_3) + \oint b_1 + \frac{pN}{4\pi}(\w_2 d\w_3 - \w_3 \bar{d}\w_2) + \frac{pN}{2\pi}(\bar{d}\w_2 \alpha_3 - d\w_3 \alpha_2) + \frac{pN}{4\pi}[\w_2 d\alpha_3 + \alpha_2d\w_3 - \w_3 \bar{d}\alpha_2 - \alpha_3\bar{d}\w_2]	\\
	=&\frac{pN}{2\pi}(\alpha_3(0)\bar{\w}_2-\alpha_2(0)\bar{\w}_3) + \oint b_1 + \frac{pN}{4\pi}(\w_2 d\w_3 - \w_3 \bar{d}\w_2) + \frac{pN}{2\pi}(\bar{d}\w_2 \alpha_3 - d\w_3 \alpha_2) + \frac{pN}{2\pi}[\alpha_2d\w_3 - \alpha_3\bar{d}\w_2]	\\
	=&\frac{pN}{2\pi}(\alpha_3(0)\bar{\w}_2-\alpha_2(0)\bar{\w}_3) + \oint b_1 + \frac{pN}{4\pi}(\w_2 d\w_3 - \w_3 \bar{d}\w_2).
	\end{split}
	\end{equation}
\end{widetext}
The calculation deserves certain explanations. In the first line after the right arrow, we have specified a particular way of coupling $A_{2,3}$ and $\alpha_{2,3}$ by choosing the lattice derivatives $d$ and $\bar{d}$. For example, at site $r$, we have:
\begin{equation}
\begin{split}
A_2(r)\alpha_3(r) \equiv& \bar{d}\w_2(r) \alpha_3(r)	\\
=& (\w_2(r)-\w_2(r-1))\alpha_3(r).
\end{split}
\end{equation}
And similarly for other terms. The first equality just expands and throws away the terms either obviously canceling each other or simply vanishing. $\alpha_{2,3}(0)$ is just the gauge transformation parameters at site $0$. In the second equality, we just pull out the integral involving $\alpha_{2,3}(0)$. In the third equality, the lattice integral by part is performed to the first and the third term in the square bracket. The boundary terms, resulting from the integral by part, double the terms in front of the loop integral. The last equality only contains the leftover terms. Using the same method, we could also find similar expressions for $b_2$ and $b_3$ related terms. Then all the terms we have are:
\begin{widetext}
	\begin{equation}
	\begin{split}
	&\frac{pN}{2\pi}n_1(\alpha_3(0)\bar{\w}_2-\alpha_2(0)\bar{\w}_3) + \oint n_1[b_1 + \frac{pN}{4\pi}(\w_2 d\w_3 - \w_3 \bar{d}\w_2)]	\\
	+&\frac{pN}{2\pi}n_2(\alpha_1(0)\bar{\w}_3-\alpha_3(0)\bar{\w}_1) + \oint n_2[b_2 + \frac{pN}{4\pi}(\w_3 \bar{d}\w_1 - \w_1 d\w_3)]	\\
	+&\frac{pN}{2\pi}n_3(\alpha_2(0)\bar{\w}_1-\alpha_1(0)\bar{\w}_2) + \oint n_3[b_3 + \frac{pN}{4\pi}(\w_1 d\w_2 - \w_2 \bar{d}\w_1)]	\\
	=& \oint n_1[b_1 + \frac{pN}{4\pi}(\w_2 d\w_3 - \w_3 \bar{d}\w_2)] + \oint n_2[b_2 + \frac{pN}{4\pi}(\w_3 \bar{d}\w_1 - \w_1 d\w_3)] + \oint n_3[b_3 + \frac{pN}{4\pi}(\w_1 d\w_2 - \w_2 \bar{d}\w_1)].
	\end{split}
	\end{equation}
\end{widetext}
Note that the terms involving $\alpha_{1,2,3}(0)$ will vanish because of the $\delta$ function constraints associated with $V_{n_1n_2n_3}$, Eq.~\eqref{eq. n1n2n3}. Therefore, we have completed the verification for the gauge invariance of the operator $V_{n_1n_2n_3}$ on lattice.

\section{Modular Matrices}
\label{app. ModMat}

In this section, we provide the modular matrices of the type III twisted $\mathbb{Z}_2^{\otimes 3}$ DW theory, as follows:

\begin{widetext}
\begin{equation}
\begin{split}
S=&\frac{1}{8}\left(
\begin{array}{cccccccccccccccccccccc}
 1 & 1 & 1 & 1 & 1 & 1 & 1 & 1 & 2 & 2 & 2 & 2 & 2 & 2 & 2 & 2 & 2 & 2 & 2 & 2 & 2 & 2 \\
 1 & 1 & 1 & 1 & 1 & 1 & 1 & 1 & -2 & 2 & 2 & -2 & 2 & 2 & -2 & -2 & 2 & -2 & -2 & 2 & -2 & -2 \\
 1 & 1 & 1 & 1 & 1 & 1 & 1 & 1 & 2 & -2 & 2 & 2 & -2 & 2 & -2 & 2 & -2 & -2 & 2 & -2 & -2 & -2 \\
 1 & 1 & 1 & 1 & 1 & 1 & 1 & 1 & 2 & 2 & -2 & 2 & 2 & -2 & 2 & -2 & -2 & 2 & -2 & -2 & -2 & -2 \\
 1 & 1 & 1 & 1 & 1 & 1 & 1 & 1 & -2 & -2 & 2 & -2 & -2 & 2 & 2 & -2 & -2 & 2 & -2 & -2 & 2 & 2 \\
 1 & 1 & 1 & 1 & 1 & 1 & 1 & 1 & -2 & 2 & -2 & -2 & 2 & -2 & -2 & 2 & -2 & -2 & 2 & -2 & 2 & 2 \\
 1 & 1 & 1 & 1 & 1 & 1 & 1 & 1 & 2 & -2 & -2 & 2 & -2 & -2 & -2 & -2 & 2 & -2 & -2 & 2 & 2 & 2 \\
 1 & 1 & 1 & 1 & 1 & 1 & 1 & 1 & -2 & -2 & -2 & -2 & -2 & -2 & 2 & 2 & 2 & 2 & 2 & 2 & -2 & -2 \\
 2 & -2 & 2 & 2 & -2 & -2 & 2 & -2 & 4 & 0 & 0 & -4 & 0 & 0 & 0 & 0 & 0 & 0 & 0 & 0 & 0 & 0 \\
 2 & 2 & -2 & 2 & -2 & 2 & -2 & -2 & 0 & 4 & 0 & 0 & -4 & 0 & 0 & 0 & 0 & 0 & 0 & 0 & 0 & 0 \\
 2 & 2 & 2 & -2 & 2 & -2 & -2 & -2 & 0 & 0 & 4 & 0 & 0 & -4 & 0 & 0 & 0 & 0 & 0 & 0 & 0 & 0 \\
 2 & -2 & 2 & 2 & -2 & -2 & 2 & -2 & -4 & 0 & 0 & 4 & 0 & 0 & 0 & 0 & 0 & 0 & 0 & 0 & 0 & 0 \\
 2 & 2 & -2 & 2 & -2 & 2 & -2 & -2 & 0 & -4 & 0 & 0 & 4 & 0 & 0 & 0 & 0 & 0 & 0 & 0 & 0 & 0 \\
 2 & 2 & 2 & -2 & 2 & -2 & -2 & -2 & 0 & 0 & -4 & 0 & 0 & 4 & 0 & 0 & 0 & 0 & 0 & 0 & 0 & 0 \\
 2 & -2 & -2 & 2 & 2 & -2 & -2 & 2 & 0 & 0 & 0 & 0 & 0 & 0 & 4 & 0 & 0 & -4 & 0 & 0 & 0 & 0 \\
 2 & -2 & 2 & -2 & -2 & 2 & -2 & 2 & 0 & 0 & 0 & 0 & 0 & 0 & 0 & 4 & 0 & 0 & -4 & 0 & 0 & 0 \\
 2 & 2 & -2 & -2 & -2 & -2 & 2 & 2 & 0 & 0 & 0 & 0 & 0 & 0 & 0 & 0 & 4 & 0 & 0 & -4 & 0 & 0 \\
 2 & -2 & -2 & 2 & 2 & -2 & -2 & 2 & 0 & 0 & 0 & 0 & 0 & 0 & -4 & 0 & 0 & 4 & 0 & 0 & 0 & 0 \\
 2 & -2 & 2 & -2 & -2 & 2 & -2 & 2 & 0 & 0 & 0 & 0 & 0 & 0 & 0 & -4 & 0 & 0 & 4 & 0 & 0 & 0 \\
 2 & 2 & -2 & -2 & -2 & -2 & 2 & 2 & 0 & 0 & 0 & 0 & 0 & 0 & 0 & 0 & -4 & 0 & 0 & 4 & 0 & 0 \\
 2 & -2 & -2 & -2 & 2 & 2 & 2 & -2 & 0 & 0 & 0 & 0 & 0 & 0 & 0 & 0 & 0 & 0 & 0 & 0 & -4 & 4 \\
 2 & -2 & -2 & -2 & 2 & 2 & 2 & -2 & 0 & 0 & 0 & 0 & 0 & 0 & 0 & 0 & 0 & 0 & 0 & 0 & 4 & -4 \\
\end{array}
\right),	\\
T=&\mathrm{diag}\left(1,1,1,1,1,1,1,1,1,1,1,-1,-1,-1,1,1,1,-1,-1,-1,-i,i
\right)
\end{split}
\end{equation}
The basis order for modular matrices is: $U_{000}$, $U_{100}$, $U_{010}$, $U_{001}$, $U_{110}$, $U_{101}$, $U_{011}$, $U_{111}$,
$(100,\alpha^1_+)$, $(010,\alpha^2_+)$, $(001,\alpha^3_+)$, 
$(100,\alpha^1_-)$, $(010,\alpha^2_-)$, $(001,\alpha^3_-)$,
$(110,\beta^3_+)$, $(101,\beta^2_+)$, $(011,\beta^1_+)$, 
$(110,\beta^3_-)$, $(101,\beta^2_-)$, $(011,\beta^1_-)$,
$(111,\gamma_+)$, $(111,\gamma_-)$.
\end{widetext}

\section{Quantum Double Calculation}
\label{app. QD}

In this section, we will show how quantum double calculation yields the same correlation in the right above section. We need to use two assumptions in our quantum double calculation: The first one is that we only choose a \textit{prime} $N$, otherwise we will encounter integer equations without an explicit solution to our knowledge; the second assumption is that we will pick a particular solution of some integer equations we encounter. That's the same thing as we pick up particular operators to calculate correlation functions. For more details and the philosophical reasons of quantum double arising from discrete gauge theories of 2 spatial dimension, please refer to Ref.~\onlinecite{propitius1995topological}.

This section will be divided into two parts: the first one we will produce the projective representation, while the second one we will calculate the correlation function via the $\mathcal{R}$ symbol.

\subsection{Projective Representation}

In this part, we will introduce the projective representation of quantum double calculation. For simplicity, we denote:
\begin{eqnarray}
\w=\exp(\frac{2\pi\ii p}{N}),
\end{eqnarray}
where $p$ is the same parameter in the Lagrangian Eq.~\eqref{lag. 2+1 ZN}. Now given by the group $\mbb{Z}_{N}^{\otimes3}$, we use the 3-cocycle as follows:
\begin{eqnarray}
\alpha(A,B,C)=\exp(\frac{2\pi\ii p}{N}A^1B^2C^3)\equiv \w\{A^1B^2C^3\},
\end{eqnarray}
where $A,B,C\in\mbb{Z}_{N}^{\otimes3}$ and more explicitly $A=(A^1,A^2,A^3)$, $B=(B^1,B^2,B^3)$ and $C=(C^1,C^2,C^3)$ for the three components of $\mbb{Z}_{N}^{\otimes3}$. Now we can define the slant product:
\begin{eqnarray}
c_A(B,C)=\frac{\alpha(A,B,C)\alpha(B,C,A)}{\alpha(B,A,C)}.
\end{eqnarray}
And the projective representation $M_A(g)~(g\in\mbb{Z}_N^{\otimes3})$ is specified by $c_A(B,C)$ as follows:
\begin{eqnarray}\label{eq.projectiverep}
M_A(B) M_A(C) = c_A(B,C) M_A(BC).
\end{eqnarray}

In particular, for $A=(n_1n_2n_3)$, we have the slant product $c_A(B,C)$ explicitly:
\begin{equation}
\begin{split}
c_A(B,C)=&\frac{\alpha(A,B,C)\alpha(B,C,A)}{\alpha(B,A,C)}	\\
=&\w\{n_1 B^2C^3 + n_3B^1C^2 - n_2B^1C^3\}.
\end{split}
\end{equation}
Then the representation of the generators in $\mbb{Z}_N^{\otimes3}$ will satisfy the following equation according to Eq.~\eqref{eq.projectiverep}:
\begin{equation}\label{eq. ZN proj rep}
\begin{split}
M_A(100)M_A(010) &= \w^{n_3}M_A(110)	\\
M_A(010)M_A(100) &= M_A(110)	\\
M_A(100)M_A(001) &= \w^{-n_2}M_A(101)	\\
M_A(001)M_A(100) &= M_A(101)	\\
M_A(010)M_A(001) &= \w^{n_1}M_A(011)	\\
M_A(001)M_A(010) &= M_A(011),
\end{split}
\end{equation}
where $A=(n_1n_2n_3)$ and we use this convention for the rest of the calculation until specified. Moreover, we will require that
\begin{equation}\label{eq. rep 2}
\begin{split}
M_A(000)&\equiv M_A(N00)=(M_A(100))^N=1	\\
M_A(000)&\equiv M_A(0N0)=(M_A(010))^N=1	\\
M_A(000)&\equiv M_A(00N)=(M_A(001))^N=1.
\end{split}
\end{equation}

One solution for the representation of the generators, which we will use to compare with field theory calculations, are:
\begin{equation}\label{eq. solu. projectiverep}
\begin{split}
M_A(100)&=\Sigma_3	\\
M_A(010)&=c_2 \Sigma_1^{n_3}	\Sigma_3^{v}	\\
M_A(001)&=c_3 \Sigma_1^{-n_2}	\Sigma_3^{y},
\end{split}
\end{equation}
where $\Sigma_3$, $\Sigma_1$ are the generalized Pauli matrix satisfying $\Sigma_3\Sigma_1=\w\Sigma_1\Sigma_3$. And $c_2$, $c_3$, $v$ and $y$ are parameters satisfying
\begin{equation}\label{eq. solu. prorep. para1}
\begin{split}
& n_1 + n_2 v + n_3 y \equiv 0 \mod{N}	\\
& c_2 = \w\{-n_3 v \frac{N-1}{2}\}	\\
& c_3 = \w\{ n_2 y \frac{N-1}{2}\}
\end{split}
\end{equation}
as a result of Eq.~\eqref{eq. ZN proj rep} and Eq.~\eqref{eq. rep 2}.

Hence, we have completed the calculation of the representations of the projective representation determined by the slant product of 3-cocycle, with several parameters.

\subsection{Braiding Statistics}

In quantum double, the braiding statistics of two particles $a$ and $b$ is generally given by the trace of squared $\mathcal{R}$ symbol, $Tr(R_{ab}R_{ba})$. For more details, please refer to the second chapter of Ref.~\onlinecite{propitius1995topological}. As a summary of the Ref.~\onlinecite{propitius1995topological}, the braiding statistics can be written as:
\begin{equation}\label{eq. mutual statistics}
\mathrm{Tr} (M_{n_1n_2n_2}(m_1m_2m_3)) \mathrm{Tr}(M_{m_1m_2m_2}(n_1n_2n_3)).
\end{equation}

Later in this section , we will verify that the correlation $\langle V_{n_1n_2n_3}(\gamma_1)V_{m_1m_2m_3}(\gamma_2) \rangle$, which is the field theoretical counterpart of braiding statistics, will appear in the solutions of Eq.~\eqref{eq. mutual statistics}, with an assumption about the solution of an integer number equation. Now we start with calculating $M_{n_1n_2n_2}(m_1m_2m_3)$:
\begin{widetext}
\begin{equation}
\begin{split}
M_{n_1n_2n_3}(m_1m_2m_3) &= \w\{-n_1m_2m_3 + n_2m_1m_3 - n_3m_1m_2\} M_{n_1n_2n_3}(m_100)M_{n_1n_2n_3}(0m_20)M_{n_1n_2n_3}(00m_3)	\\
&=\w\{-n_1m_2m_3 + n_2m_1m_3 - n_3m_1m_2\}
M_{n_1n_2n_3}(100)^{m_1}M_{n_1n_2n_3}(010)^{m_2}M_{n_1n_2n_3}(001)^{m_3}	\\
&=\w\{-n_1m_2m_3 + n_2m_1m_3 - n_3m_1m_2\} (\Sigma_3)^{m_1} (c_2\Sigma_1^{n_3}\Sigma_3^{v})^{m_2} (c_3\Sigma_1^{-n_2}\Sigma_3^{y})^{m_3}.
\end{split}
\end{equation}
\end{widetext}
Notice that the trace of $M_{n_1n_2n_3}(m_1m_2m_3)$ will vanish except when:
\begin{equation}\label{eq. solu. prorep. para2}
\begin{split}
& m_1 + m_2 v + m_3 y \equiv 0 \mod{N}	\\
& n_3m_2 \equiv n_2m_3 \mod{N}.
\end{split}
\end{equation}
If we select other ways of representing the solutions of the projective representations, we would also yield:
\begin{equation}\label{eq. solu. prorep. para3}
\begin{split}
& n_1m_2 \equiv n_2m_1 \mod{N}	\\
& n_1m_3 \equiv n_3m_1 \mod{N}.
\end{split}
\end{equation}

Now the matrix $M_{n_1n_2n_3}(m_1m_2m_3)$ is proportional to identity with the coefficient:
\begin{widetext}
\begin{equation}\label{eq. phase1}
\begin{split}
& M_{n_1n_2n_3}(m_1m_2m_3)	\\
=&\w\{-n_1m_2m_3 + n_2m_1m_3 - n_3m_1m_2\} c_2^{m_2} c_3^{m_3}  \w\{n_3v\frac{m_2(m_2-1)}{2}-n_2y\frac{m_3(m_3-1)}{2} - n_3m_2^2v\}	\\
=& \w\{-n_1m_2m_3 + \frac{n_2m_3}{2}(2m_1 + Ny - m_3y) - \frac{n_3m_2}{2}(Nv + m_2v + 2m_1)\}.
\end{split}
\end{equation}
\end{widetext}

Symmetrically, we can yield that:
\begin{equation}\label{eq. phase2}
\begin{split}
&M_{m_1m_2m_3}(n_1n_2n_3)	\\
=& \w\{-m_1n_2n_3 + \frac{m_2n_3}{2}(2n_1 + N\tilde{y} - n_3\tilde{y}) 	\\
& - \frac{m_3n_2}{2}(N\tilde{v} + n_2\tilde{v} + 2n_1)\},
\end{split}
\end{equation}
which parameters satisfying:
\begin{equation}
\begin{split}
& n_1 + n_2 \tilde{v} + n_3 \tilde{y} \equiv 0	\mod{N}	\\
& m_1 + m_2 \tilde{v} + m_3 \tilde{y} \equiv 0	\mod{N}.
\end{split}
\end{equation}

Now we make the following assumptions about the solution, or we select a particular solution for the parameters in order to compare a particular correlation function in the previous section:
\begin{equation}
\begin{split}
& m_i = m t_i, n_i = n t_i,~i=1,2,3	\\
& y = \tilde{y},~v = \tilde{v}\;.
\end{split}
\end{equation}
And $m$ and $n$ are mutual prime numbers. Therefore, we can conclude that $t_1 + t_2 v + t_3 y \equiv kN, k\in\mathcal{Z}$.

Finally the mutual statistics is:
\begin{equation}
\begin{split}
&\mathrm{Tr}(M_{n_1n_2n_3}(m_1m_2m_3)) \mathrm{Tr}(M_{m_1m_2m_3}(n_1n_2n_3))	\\
=& N^2 \w\{-\frac{nm^2 + n^2m}{2}t_1t_2t_3 - \frac{mn}{2} t_2t_2(m+n)kN\}.
\end{split}
\end{equation}
Note that $mn(m+n)$ is always an even number. Therefore the $\frac{mn}{2} t_2t_2(m+n)kN$ is a multiple of $N$. Therefore the second term is only a trivial phase. Hence, we have the final result to be:
\begin{equation}
\begin{split}
&\mathrm{Tr}\left( M_{n_1n_2n_3}(m_1m_2m_3)\right) \mathrm{Tr}\left(M_{m_1m_2m_3}(n_1n_2n_3) \right)	\\
=& N^2 \w\{-\frac{nm^2 + n^2m}{2}t_1t_2t_3\}	\\
=& N^2 \exp\{-\frac{2\pi\ii p}{N}\frac{nm^2 + n^2m}{2}t_1t_2t_3\}.
\end{split}
\end{equation}
Remember that the condition to prevent the trace from vanishing is that $n_i$ and $m_i$ are proportional to each other.
Until now, we have complete the calculation of the braiding statistics and it is the same value as its field theoretical counterpart $\langle V_{n_1n_2n_3}(\gamma_1)V_{m_1m_2m_3}(\gamma_2) \rangle$, which we have calculated in the previous section.

\bibliography{NonFlatDW}

\end{document}